\documentclass[%
reprint,
 amsmath,amssymb,
 aps,
]{revtex4}

\usepackage{graphicx}
\usepackage{dcolumn}
\usepackage{bm}
\usepackage{appendix}
\usepackage[dvipsnames]{xcolor}
\DeclareMathOperator{\Tr}{Tr}
\begin{document}

\title{Modelling Chromatic Emittance Growth in Staged Plasma Wakefield Acceleration to 1 TeV using Nonlinear Transfer Matrices}

\author{Alec G. R. Thomas}
\affiliation{G\'erard Mourou Center for Ultrafast Optical Sciences
}
\affiliation{Department of Nuclear Engineering and Radiological Sciences, University of Michigan, Ann Arbor, MI 48109, USA}
\author{Daniel Seipt}

\affiliation{Helmholtz Institut Jena, Fr\"obelstieg 3, 07743 Jena, Germany}
\affiliation{GSI Helmholtzzentrum f\"ur Schwerionenforschung GmbH, Planckstrasse 1, 64291 Darmstadt, Germany}
\date{\today}

\begin{abstract}
A framework for integrating transfer matrices with particle-in-cell simulations is developed for TeV staging of plasma wakefield accelerators. Using nonlinear transfer matrices in terms up to ninth order in normalized energy spread $\sqrt{\langle\delta\gamma^2\rangle}$ and deriving a compact expression for the chromatic emittance growth in terms of the nonlinear matrix, plasma wakefield  accelerating stages simulated using the three-dimensional particle-in-cell framework OSIRIS 4.0 were combined to model acceleration of an electron beam from 10 GeV to 1 TeV in 85 plasma stages of meter scale-length with long density ramps and connected by simple focusing lenses. In this calculation, we find that for {initial} relative energy spreads below $10^{-3}${, energy-spread growth  below $10^{-5}$ of the energy gain per stage} and normalized emittance below mm-mrad, the chromatic emittance growth can be minimal. The technique developed here may be useful for plasma collider design, and {potentially} could be expanded to encompass non-linear wake structures and include other degrees of freedom such as lepton spin.
\end{abstract}

\maketitle

\section{\label{sec:level1}Introduction}
Laser and beam driven plasma wakefield acceleration are promising approaches for accelerating leptons to high energy \cite{NJP_Roadmap} and plans for a plasma based accelerator facility are at a mature stage \cite{eupraxia}. For collider applications, energies in excess of 100 GeV will be required and it is likely that multiple plasma stages are required \cite{Lindstrom_2021}. There has been a lot of work in understanding transport between stages, through experiments \cite{Steinke_Nature_2016} and simulations/theory \cite{Antici_JAP_2012,Migliorati_PRZ_2013,Xu_PRL_2016,Benedetti_PRAB_2017}. In particular, significant attention has been paid to chromatic emittance growth through mismatched beams \cite{Mehrling_PRZ_2012} and misalignment {\cite{Cheshkov_PRZ_2000,Chiu_PRZ_2000,Thevenet_PRAB_2019}}. To improve the matching, adiabatic matching using density ramps has been studied \cite{Floettmann_PRZ_2014,Dornmair_PRZ_2015,Litos_PTRSA_2019,Ariniello_PRZ_2019,Zhao_PRZ_2020} as well as other beam transport components based on plasma elements \cite{Manahan_NComms_2017,Lindstrom_PRL_2018,DArcy_PRL_2019,Pousa_PRL_2019}. 

In conventional accelerators beam transport is a mature subject \cite{Wiedemann_book,Wolski_book}, in particular the use of transfer matrices to describe the particle dynamics. There is interest in finding fast particle tracing methods for plasma accelerators \cite{Ferran_Pousa_2019}. {Analytic solutions for wakefields have been used as the basis for developing transfer matrices for studying staging of plasma accelerators \cite{Cheshkov_PRZ_2000,Chiu_PRZ_2000}.} In this paper, we show how transfer matrices for plasma accelerators can be constructed from the fields calculated in self-consistent particle-in-cell simulations. Having such a framework allows integration of plasma elements simulated with particle-in-cell codes with conventional accelerator design codes/techniques. This method is not a replacement for full-scale simulations, as feedback of the beam on the wakefields cannot be included. But full scale particle-in-cell simulations of a multi-stage plasma accelerator are  computationally expensive, so having an approach that may allow rapid design of complex lattices involving plasma accelerating stages and other elements, such as plasma optics, should prove useful.

Analytic solutions can also be used to model the particle transport, i.e. Wentzel-Kramers-Brillouin (WKB) solutions for the betatron oscillations, but there are limitations. First, the density ramps at the beginning and end of the accelerator have been determined to be crucial for staging \cite{Xu_PRL_2016,Benedetti_PRAB_2017} but, especially at high energies, the ramp length can become comparable to the betatron wavelength and so the WKB approximation breaks down at the plasma-vacuum interface. Further, particularly in laser driven wakefields, the evolution of the pulse and wakefield could be complex for efficient acceleration in the nonlinear regime and hard to capture without resorting to full scale simulation. With the approach described here, a full scale simulation is required to be performed, but once performed, the same simulation may be used to study different beam phasespaces rapidly and combined with different elements to build an accelerating lattice.

\begin{figure*}
    \centering
    \includegraphics[width = \textwidth]{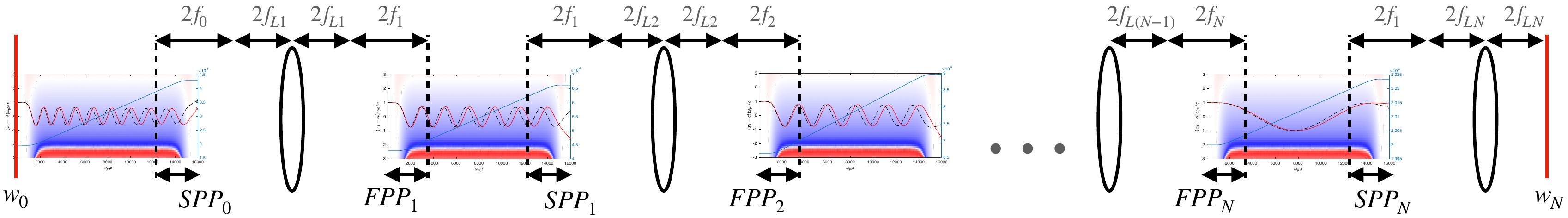}
    \caption{Schematic showing $N$ stage plasma accelerator, with (chromatic) focusing lenses between plasma accelerating stages with $2f$ focusing throughout. {$f_s$ is the plasma stage focal length and $f_{Ls}$ is the beam focusing optic focal-length (at the design energy) of the $s$th accelerating stage.} $FPP_s$ and $SPP_s$ are the primary and secondary principal planes of the accelerating stage respectively.}
    \label{fig:Lattice}
\end{figure*}

The paper proceeds in the following manner. Section \ref{sec:level2} lays out the framework for  transfer matrices $\mathcal{M}$ that are nonlinear in the beam energy (spread). Section \ref{sec:level3} derives the chromatic emittance growth from the nonlinear transfer matrix by defining an extended beam matrix $\Sigma$,  such that the emittance growth can be calculated using the expression 
\begin{equation}
    \epsilon_N = \sqrt{\det\left(\mathcal{P}^T\mathcal{M}\Sigma\mathcal{M}^T\mathcal{P}\right)}\;,
\end{equation}
where $\mathcal{P}$ is a projector. 
Section \ref{sec:level4} calculates the nonlinear transfer matrices for other simple elements, i.e. drift space and simple focusing lens, for demonstration of combining the plasma accelerator simulations with other elements. Section \ref{sec:10gev} describes a three-dimensional particle-in-cell simulation of a meter-scale beam-driven plasma-wakefield accelerator and the construction of a set of transfer matrices through the stages. Finally, Section \ref{sec:focusing} outlines a design for a simple lattice comprising `cells' of a plasma accelerating stage, two drift spaces and a simple (thin) lens accelerating a beam of particles from 10 GeV to 1 TeV --- as shown in the schematic in Fig.~\ref{fig:Lattice} --- and calculates the resulting chromatic emittance growth as a function of initial transverse emittance and beam energy spread.

\section{\label{sec:level2} Linear transfer matrices for plasma accelerators}
Assuming a coordinate system $x,y,z$, we can build transfer matrices from particle-in-cell simulations performed in a window moving at the speed of light in the $z$ direction, as is typical, by assuming that the beam is ultrarelativistic, $1 - \beta_z \lll1$, where its normalized velocity is $\beta_z = v_z/c$. This assumption means that the beam remains at approximately constant phase, $z - ct$, and therefore experiences fields at a fixed grid position in the simulation box, i.e. time dependent fields only. By making use of a paraxial approximation, the field gradients on the axis at that fixed grid position integrated over time are the only information required to build the matrix describing the transport of a beam with a given energy (the ``design energy'') through the full plasma accelerator.

\subsection{Basic transfer matrix}
We start with the equations of motion for a charged particle with charge $q$ and mass $m$
{in external fields $\vec{E}$ and $\vec{B}$,}
\begin{equation} \label{eq:dxdt}
\frac{dx}{dt} =  \frac{u_x}{\gamma}\;,\quad\frac{dy}{dt} =  \frac{u_y}{\gamma}\;,
\end{equation}
\begin{equation}
\frac{du_x}{dt} \simeq \frac{q}{m}\left(E_x -cB_y\right)\;,
\quad\frac{du_y}{dt} \simeq \frac{q}{m}\left(E_y +cB_x\right)\;,\label{momeqs}
\end{equation}
and
\begin{equation}
\frac{d\gamma}{dt} = \frac{qE_z}{mc}\;,
\end{equation}
where $u_l = \gamma v_l$ is the proper velocity, with $l$ a Cartesian component ($l = x,y$).

Under the paraxial approximation, we may expand the $E_x$, $E_y$, $B_x$ and $B_y$ fields as a Taylor series in $x$ and $y$ about the axis; 
\begin{equation}
F_l(x,y,z) = F_l(0,0,z) + x\frac{\partial F_l}{\partial x}(0,0,z) + y\frac{\partial F_l}{\partial y}(0,0,z) + \dots
\end{equation}
Where $F$ is a field ($F =E,B$). Hence, Eqns.~\eqref{momeqs} can be expressed as 
\begin{equation} \label{eq:dudt}
    \frac{du_x}{dt} \simeq -\alpha^2_{xx}x-\alpha^2_{xy} y\;,\quad \frac{du_y}{dt} \simeq -\alpha^2_{yx}x-\alpha^2_{yy}y\;,
\end{equation}
where
\begin{equation}
\alpha^2_{kl} =  - \frac{q}{m}\left.\frac{\partial}{\partial k} \left(E_l -\varepsilon_{lp} cB_p\right)\right|_{x=0,y=0}\;,
\end{equation}
with $\varepsilon_{lp}$ the Levi-Civita symbol and using Einstein summation convention.

We want to solve the equation of motion piece-wise in the form of a series of matrix solutions for each timestep in the simulation that may be combined to form a single matrix for propagation of a charged particle beam through a whole simulated plasma component (accelerating stage, lens etc.). The matrices will need to be sufficiently accurate in betatron phase to consider a  large number of oscillations and it should be symplectic to conserve beam emittance, for a beam with all particles having the same energy.

We further make the following  assumptions/approximations;
\begin{itemize}
    \item The beam energy $\gamma mc^2$ slowly varies compared to the time step size. This means that in the transverse equation of motion it is assumed constant over a timestep, but the beam energy is increased each step by ${qE_z\Delta t}/{mc}$, i.e. $$\gamma^n  \simeq \gamma_{0}+ \sum_{n^\prime=0}^n\frac{q{E_{z}}^{n^\prime}\Delta t}{mc}\;,$$ where the superscript $n$ denotes the time step. 
    \item The timestep is small compared to the plasma (laser) period, and is therefore extremely small compared to the betatron frequency. It is therefore not necessary to use the usual accelerator physics, e.g. $C_x = \cos(\alpha_{xx}\Delta t/\sqrt{\gamma})$ and  $S_x = \sin(\alpha_{xx}\Delta t/\sqrt{\gamma}) / \alpha_{xx}\sqrt{\gamma}$, solutions to the harmonic oscillator equation. This simplifies expanding the transfer matrix to arbitrarily higher order perturbations. 
    \item The force is curl free, i.e. conservative. This means that $\alpha_{xy}^2 = \alpha_{yx}^2$.
\end{itemize}

We can write the equations of motion in matrix form as 
$$
\frac{dw}{dt} = A^nw\;,
$$
where, as before, the superscript $n$ denotes the time step, 
$$
A^n= \begin{bmatrix}
0 & \frac{1}{\gamma^n} & 0 & 0\\
- [\alpha_{xx}^2]^n & 0 & -[\alpha_{xy}^2]^n& 0\\
0 & 0 & 0 &\frac{1}{\gamma^n}\\
- [\alpha_{yx}^2]^n& 0 & - [\alpha_{yy}^2]^n & 0
\end{bmatrix}\quad{\rm and}\quad w = \begin{bmatrix}
x^{n} \\ u_{x}^{n}  \\ y^{n} \\ u_{y}^{n}  \end{bmatrix}\;.
$$
The solution to this equation over a time step $\Delta t$ is
$$
w^{n+1} = e^{A^n\Delta t}w^{n}\;.
$$
 If we truncate the series representing the matrix exponential at e.g. second order, the solution is not symplectic. We solve this issue by splitting the matrix  $A^n$ into two matrices such that $A^n = A_1^n + A_2^n$ \cite{BuonoLopez},  where
 $$
A^n_1= \begin{bmatrix}
0 & \frac{1}{\gamma^n} & 0 & 0\\
0 & 0 & 0& 0\\
0 & 0 & 0 &\frac{1}{\gamma^n}\\
0& 0 &  0 & 0
\end{bmatrix}\;,\quad 
A^n_2= \begin{bmatrix}
0 & 0& 0 & 0\\
- [\alpha_{xx}^2]^n & 0 & -[\alpha_{xy}^2]^n& 0\\
0 & 0 & 0 &0\\
- [\alpha_{yx}^2]^n& 0 & - [\alpha_{yy}^2]^n & 0
\end{bmatrix}
\;.
$$

 From the Baker–Campbell–Hausdorff relation, $e^{A^n_1\Delta t}e^{A^n_2\Delta t} = e^{(A^n_1+A^n_2)\Delta t + \frac{1}{2}\Delta t^2[A^n_1,A^n_2] +\dots}$, i.e. $e^{A^n_1\Delta t}e^{A^n_2\Delta t}$ is an approximation of the exact solution to (at least) second-order accuracy in $\Delta t$. 

 For a nonsingular, skew-symmetric matrix $\Omega$, it can be shown that for $X \in \{e^{A_1^n\Delta t}, e^{A_2^n\Delta t}\}$, $X^T \Omega X = \Omega$ and hence $X$ is symplectic, 
 provided that $[\alpha_{xy}^2]^n =  [\alpha_{yx}^2]^n$, which is the case for a conservative force, $\nabla\times \vec{F} = 0$.  As $A_i^n$ are nilpotent, their matrix exponentials can be calculated exactly and combined to give a symplectic, second order accurate solution to the equations of motion over a timestep $w^{n+1} = M^nw^n$, using the matrix $M^n = e^{A^n_1\Delta t}e^{A^n_2\Delta t} = e^{A^n\Delta t + \mathcal{O}(\Delta t^2)}$, i.e. 
\begin{equation}
\begin{bmatrix}
x^{n+1} \\ u_{x}^{n+1}  \\ y^{n+1} \\ u_{y}^{n+1}  \end{bmatrix} 
= \begin{bmatrix}
1-\frac{[\alpha_{xx}^2]^n\Delta t^2}{\gamma^n} & \frac{\Delta t}{\gamma^n} & -\frac{[\alpha_{xy}^2]^n\Delta t^2}{\gamma^n} & 0\\
- [\alpha_{xx}^2]^n\Delta t & 1 & -[\alpha_{xy}^2]^n\Delta t & 0\\
-\frac{[\alpha_{yx}^2]^n\Delta t^2}{\gamma^n} & 0 & 1-\frac{[\alpha_{yy}^2]^n\Delta t^2}{\gamma^n} &\frac{\Delta t}{\gamma^n}\\
- [\alpha_{yx}^2]^n\Delta t & 0 & - [\alpha_{yy}^2]^n\Delta t & 1
\end{bmatrix}
\cdot
\begin{bmatrix}
x^{n} \\ u_{x}^{n}  \\ y^{n} \\ u_{y}^{n}  \end{bmatrix}
\;,\label{eqn_matrix}
\end{equation}
where, in particular, $\det M^n=1$.
We form the full transfer matrix by calculating each matrix corresponding to the transformation of the coordinates over a time-step and then combining these  to form a single matrix describing the propagation through the plasma element. To obtain $\partial E_x/\partial x|_{x=0,y=0}$ etc.  from a simulation, the numerical gradient can be taken near the axis. 
We introduce another index $j$ to the matrix, because the equation of motion is solved at a particular phase $\xi_j = z_j-ct$ corresponding to the position at grid point $j$, so that $w_j^{n+1} = M_j^nw_j^n$. We can write down the transfer matrix through the whole system at a particular phase $M_j$ using the time ordered product 
\begin{equation}
    M_j(\gamma_0) = \prod_{n=0}^{N_t} M_j^n(\gamma_j^n)\;,
\end{equation}
where the dependence on the particle initial energy is explicitly shown, so that the particle coordinates are transformed through the full plasma element as
\begin{equation}
    w = M_j(\gamma_0)\cdot w_0\;.
\end{equation}
The set of matrices $M_j$ are functions of the initial particle energy, and hence need recalculating for each energy of particle passing through the plasma. The transfer matrix is therefore calculated for a ``design energy'' for particles passing through each plasma element, and then arbitrary transverse distributions may be then studied using the resulting matrix. The relative phase error is second order,  as shown in Appendix~\ref{phaseerror}, and therefore negligible when using the high-resolution fields from particle-in-cell simulations. 

\subsection{Energy spread considerations}
For plasma accelerators, one consideration of interest is the effect of the beam energy spread. Because of the requirement to include the beam energy in the calculation of the transfer matrix, we need to find a different way to approach the effect of particles having different energies without resorting to brute force calculation of $M_j$ for every value of initial beam energy $\gamma_0$. 

To do this, we consider a perturbation to the initial particle energy, $\gamma = \gamma_0 + \delta\gamma$. The usual approach in standard accelerator theory \cite{Wiedemann_book},   similar to that developed in \cite{Brown_APP_1982}, is to consider the perturbed solution using a Green's function approach to the homogeneous equation
$$
G(\tau,\tau^\prime) = S(\tau)C(\tau^\prime) - S(\tau^\prime)C(\tau)\;,
$$
and then adding in the resulting  terms into a new nonlinear matrix. However, for the transfer matrix given in Eqn.~\eqref{eqn_matrix}, derivatives of $M$ are proportional to successive powers of $1/\gamma$, which simplifies the approach and allows us to easily expand to arbitrary order in the perturbation $\delta\gamma$.

For compactness of notation in this section, we drop the $j$ and $n$ indices on the quantities $M_j^n$ etc. in this section.
For a particle with energy $\delta\gamma$  from the design energy, its transfer matrix is
$$M(\gamma+\delta\gamma) = e^{\tfrac{\gamma}{\gamma+\delta\gamma}A_1\Delta t}e^{A_2\Delta t}\;.$$
Using $\frac{\gamma}{\gamma+\delta\gamma} = 1 - \frac{\delta \gamma}{\gamma+\delta\gamma}$, we can express
\begin{eqnarray}
M(\gamma+\delta\gamma)
&=& M(\gamma) - \frac{\delta\gamma}{\gamma+\delta\gamma}M_D(\gamma)\label{mdeqn}
\end{eqnarray}
where 
$$
M_D =  A_1\Delta t e^{A_2\Delta t}\;.
$$
 For the specific case of the $4\times 4$ transfer matrix given in Eqn.~\eqref{eqn_matrix},
\begin{equation}
    M_D = \begin{bmatrix}
-\frac{\alpha^2_{xx}\Delta t^2}{\gamma} & \frac{\Delta t}{\gamma} & -\frac{\alpha^2_{xy}\Delta t^2}{\gamma} & 0\\
0 & 0 & 0 & 0\\
-\frac{\alpha^2_{yx}\Delta t^2}{\gamma} & 0 & -\frac{\alpha^2_{yy}\Delta t^2}{\gamma} &\frac{\Delta t}{\gamma}\\
0 & 0 & 0 & 0
\end{bmatrix} .
\end{equation}

The second term on the right-hand-side of Eqn.~\eqref{mdeqn} can be expanded as a Taylor series in $\delta\gamma/\gamma$;
\begin{equation}
    M(\gamma+ \delta\gamma) = M(\gamma) + \sum_{p=1}^{\infty}\left(-\frac{\delta\gamma}{\gamma}\right)^p M_D(\gamma)\label{generator}
\end{equation}

We may use series to expand the transfer matrix into a nonlinear transfer matrix that includes perturbation terms in  $\delta\gamma x$, $\delta\gamma^2 x$, $\dots$, $\delta\gamma u_x$, $\delta\gamma^2 u_x$, $\dots$, $\delta\gamma y$, $\delta\gamma^2 y$ etc., resulting in a matrix equation of the form
$$
w_\delta = \mathcal{M}w_{\delta0}\;,
$$
where $\mathcal{M}$ is the nonlinear transfer matrix and 
 \begin{equation}
         w_\delta = \begin{bmatrix}1 \\\delta\gamma \\\delta\gamma^2 \\\delta\gamma^3 \\\delta\gamma^4 \\\delta\gamma^5 \\\delta\gamma^6 \\ \vdots\end{bmatrix} \otimes w\;,\label{eq:wdelta}
 \end{equation}
where $\otimes$ denotes the Kronecker matrix product.  The matrix can be expanded to arbitrarily high terms in $\delta\gamma$ (Note that we expand in powers of $\delta\gamma$ rather than $\delta = \delta\gamma/\gamma$ because even though the equations would be more compact, $\delta$ is not a constant as the particle is in general accelerated in energy.) For staged plasma accelerators we may wish to go to a high number of orders because of the relatively large energy spread and acceleration over many betatron periods (see Appendix \ref{nonlindetails}). We may generate  the elements in the expanded matrix using Eqn.~\eqref{generator} through the relation
\begin{equation}
    \delta\gamma^p w = M(\gamma+ \delta\gamma) \delta\gamma^p w_0\;.
\end{equation}
Including up to $m$ terms in the expansion, each row $p$ of the resulting $ n(m+1)\times n(m+1)$ transfer matrix (where $n=2$ or $n=4$ depending on whether $w$ is the $2\times1$ or $4\times1$ vector describing the $x,u_x$ or $x,u_x,y,u_y$ phasespace coordinates for the particle respectively) corresponding to the transformation of $\delta\gamma^p w$ will comprise the series of terms in the expansion of $M(\gamma+ \delta\gamma)$ up to $m^\prime = m - p$. The resulting matrix can be expressed as
\begin{equation}
    \mathcal{M} = \mathbb{I}_{m+1}\otimes M + {\Gamma}\otimes M_D\;,
\end{equation}
where $\mathbb{I}_{m+1}$ is the $(m+1)\times (m+1)$ identity matrix and ${\Gamma}$ is the $(m+1)\times (m+1)$ strictly upper triangular matrix with elements at row $a$ and column $b$, (where $a,b= 0,1,\dots, m$) defined as
\begin{equation}
    {{\Gamma}}_{a,b} =  \begin{cases} \left(-\frac{1}{\gamma}\right)^{{b-a}} & b>a\\
    0 & {\rm otherwise}\;
    \end{cases}\label{gammafactor}
\end{equation}
Explicit forms of this matrix and verification of this approach are given in Appendix~\ref{nonlindetails}. As shown in this Appendix, the number of terms required for an accurate solution may be estimated from the requirement that
\begin{equation}
\left|\frac{\delta\gamma}{2}\int_0^{\psi_0}\frac{d\psi}{\gamma}\right|^m\ll m!\label{acc_check_notappendix}
\end{equation} 
for the highest order $m$ in the expansion, where $d\psi = \alpha_{kl} dt / \sqrt{\gamma}$ is the differential (betatron) phase.

As before, we may compose the transfer matrix for propagation through the whole plasma accelerator section at a particular wake phase $\mathcal{M}_j$ using the time ordered product
\begin{equation}
    \mathcal{M}_j(\gamma_0) = \prod_{n=0}^{N_t} \mathcal{M}_j^n(\gamma_j^n)\;,
\end{equation}

\section{\label{sec:level3}Chromatic emittance growth}
\begin{figure}
    \centering
    \includegraphics[width = 0.3\textwidth]{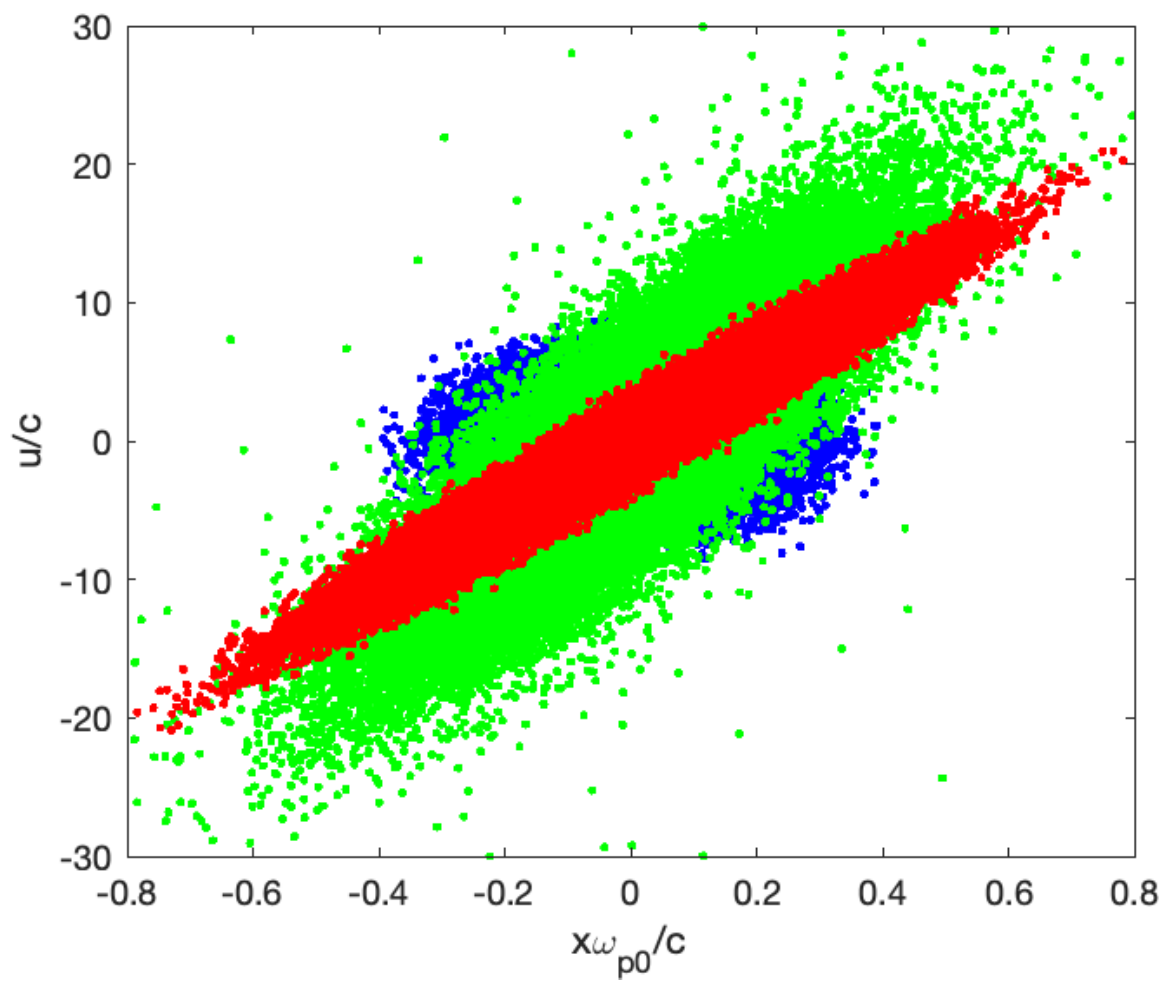}
    \caption{Phase space coordinates of 100,000 particles for an initial beam matrix shown in blue. Red indicates particle coordinates having propagated through a transfer matrix and green show the coordinates for the corresponding nonlinear matrix with randomly sampled $\delta\gamma$.  }
    \label{emittgrowth}
\end{figure}

As is customary in accelerator physics \cite{Wiedemann_book}, we may consider the $n$-dimensional phase-space ellipse defined in terms of a $n\times n$ dimensional beam matrix $\sigma$ that obeys
\begin{equation}
    w^T\sigma^{-1} w = 1\;,
\end{equation}
with the volume of the $n$-dimensional phase-space ellipse being proportional to the product of the beam transverse \emph{normalized} emittances. Assuming that at each phase $\xi_j$ the beam matrix is initially given by $\sigma_{0j}$, the beam matrix transforms through the plasma element according to 
\begin{equation}
    \sigma_j = M_j\sigma_{0j}(M_j)^T\;.\label{beammtrx}
\end{equation}
The transformation of phase space ellipse can therefore be easily calculated from the transfer matrix $M_j$.
It is well known that one of the challenges with plasma accelerators is that energy spread can lead to normalized emittance growth through betatron phase mixing \cite{Mehrling_PRZ_2012}. For illustrative purposes, in Fig.~\ref{emittgrowth}, the normalized emittance growth due to this phase mixing is shown for propagation of a large number of particles through a (nonlinear) matrix. The figure shows the phase-space coordinates of 100,000 particles sampled from an initial beam matrix $\sigma_0 = \begin{bmatrix}
    0.01& 0\\ 0& 5
    \end{bmatrix}$, shown in blue. Red indicates particle coordinates  having propagated through a transfer matrix  $M = \begin{bmatrix}
    1.1225  &0.0680\\
   19.9648   & 2.1000
    \end{bmatrix}$ (which is the matrix corresponding to the first accelerating stage calculated in the later section) and green show the tracks for the corresponding nonlinear matrix (not explicitly given for brevity), with energy displacements $\delta\gamma$  randomly sampled from a normal distribution with width $\sqrt{\langle\delta\gamma^2\rangle} =1560$. The normalized emittances  calculated from the particle distribution, using the expression
   \begin{equation}
    \epsilon_N = \sqrt{\langle x^2\rangle\langle u^2\rangle - \langle xu \rangle^2}\;,\label{emit:simps}
    \end{equation}
    (in contrast to the often used $\epsilon_N = \gamma\beta\epsilon$, where $\epsilon$ is the trace-space emittance, which are equivalent for certain distributions \cite{Floettmann_PRZ_2003}) are initially $\epsilon_N = 0.22315$ and then after passing through the matrix, $\epsilon_N = 0.22315$ for the  distribution represented by the red dots corresponding to no energy spread, and $\epsilon_N = 0.59663$ for the distribution represented by the green dots, corresponding to a distribution with energy spread 
    $\sqrt{\langle\delta\gamma^2\rangle} = 1560$. 

We now derive the chromatic emittance growth from the nonlinear matrix $\mathcal{M}$.  {For brevity, we drop the $j$ index in the following, but note that this growth is calculated for a distribution at a given (discretized) wake phase $\xi_j$ with finite (slice) energy spread.} 
From Eqn.~\eqref{beammtrx}, after passing through  transfer matrix $M$, the beam matrix $\sigma$ for a particle with energy $\gamma+\delta\gamma$ will transform as
\begin{equation}
    \sigma = M(\gamma+\delta\gamma)\sigma_0(\gamma +\delta\gamma) [M(\gamma+\delta\gamma)]^T\;.
\end{equation}
We can calculate this transformation using the perturbative method by noting that the first two rows of the matrix $\mathcal{M}$ multiplied by successive powers of $\delta\gamma$ is equivalent to the expansion given in Eqn.~\eqref{generator} --- up to the highest order term included in the matrix --- and so we may express 
\begin{equation}
     M(\gamma+\delta\gamma) \simeq \mathcal{P}^T\mathcal{M} \mathcal{G}\;,
\end{equation}
where
$$
\mathcal{P} = 
\begin{bmatrix}1\\0\\0\\0\\ \vdots\end{bmatrix}\otimes\mathbb{I}_n 
$$
is a projector from the $w_\delta$ to the $w$ subspace, i.e. it can be used to extract the first $n$ columns (or rows if transposed) 
of the matrix $\mathcal{M}$, 
\begin{equation}
 \mathcal{G} =
\begin{bmatrix}1\\ \delta\gamma\\ \delta\gamma^2\\ \delta\gamma^3\\ \vdots\end{bmatrix}\otimes\mathbb{I}_n\;, \label{Gdef}
\end{equation}
and  $\mathbb{I}_n$ is the $n\times n$ (i.e. $2\times2$ or $4\times4$) identity matrix. (Using the definition in Eqn.~\eqref{Gdef}, we can express $w_\delta = \mathcal{G}w$). Hence, defining $M_\delta \equiv M(\gamma+\delta\gamma) \equiv \mathcal{P}^T\mathcal{M} \mathcal{G}$, for a beam of energy $\gamma+\delta\gamma$, the beam matrix $\sigma$ transforms as
\begin{equation}
   \sigma(\delta\gamma) =  M_\delta \sigma_0 [M_\delta]^T\;,\label{eq:33}
\end{equation}
assuming that the initial beam matrix is identical for all particle energies, $\sigma_0$.

The emittance growth due to the energy spread $\delta\gamma$ can therefore be calculated from the chromatic variation in the beam matrix. Assuming the beam energy distribution about the mean energy $\gamma$, $\rho(\delta\gamma)$, is described by a normal distribution with energy spread $\sqrt{\langle\delta\gamma^2\rangle}$ defined as
\begin{equation}
    \rho(\delta\gamma) = C \exp\left(-\frac{\delta\gamma^2}{2\langle\delta\gamma^2\rangle}\right)\;,\label{eq:normal}
\end{equation}
where $C$ is a normalizing constant, 
then the  beam matrix averaged over $\delta\gamma$  is 
\begin{equation}
    \langle\sigma\rangle = \int_{-\infty}^\infty \rho(\delta\gamma) \sigma(\delta\gamma) d(\delta\gamma) = \int_{-\infty}^\infty \rho(\delta\gamma) M_\delta \sigma_0 [M_\delta]^T d(\delta\gamma)\;.\label{eq:avesig}
\end{equation}
Technically, the distribution in energy cannot be gaussian since $\gamma \pm \delta\gamma \geq 1$, hence the lower limit in the $\delta \gamma$ integral cannot be $-\infty$. However, the corresponding longitudinal momentum distribution can be defined in the range $(-\infty,\infty)$ with a gaussian distribution. To within the paraxial approximation and provided 
$\sqrt{\langle\delta\gamma^2\rangle}/\gamma_0\ll1$, 
these distributions are equivalent.

$\sigma(\delta\gamma)$ can be expressed as a power series in $\delta\gamma$ up to order $2m$ (since $M_\delta$ is applied left and right to $\sigma_0$), where $m$ is the maximum order in the expansion $M_\delta$,
$$
\sigma(\delta\gamma) = \sum_{p=0}^{2m}{a_p}\delta\gamma^p\;,
$$
where $a_p$ is the $p$th term in the power series. For example, 
$$
{a_1} = -\frac{1}{\gamma}\left(M_D\sigma_0 M^T +M\sigma_0 M_D^T\right)\;.
$$
In this case, Eqn.~\eqref{eq:avesig} becomes
\begin{equation}
    \langle\sigma\rangle = \sum_{p = 0}^{p=2m}{a}_p\int_{-\infty}^\infty \rho(\delta\gamma) \delta\gamma^{p} d(\delta\gamma) \;.
\end{equation}
For the normal distribution given in Eqn.~\eqref{eq:normal} the odd terms in $(\delta\gamma)^{p}$ integrate to zero and the even terms yield 
\begin{equation}
    \langle\sigma\rangle = \sum_{p = 0,\mathrm{even}}^{p=2m}\frac{2^{p/2}}{\sqrt{\pi}}\Gamma\left(\frac{p+1}{2}\right)\langle\delta\gamma^2\rangle^{p/2} {a}_p \;,
\end{equation}
where $\Gamma(z)$ is the gamma function, which can be simplified to
\begin{equation}
     \langle\sigma\rangle = \sum_{p=0}^{m} \, (2p-1)!! \, \langle\delta\gamma^2\rangle^{p} \, {a}_{2p} \;,
\end{equation}
where $x!!$ indicates the double factorial of $x$.

To calculate $\langle\sigma\rangle$ in a convenient way, we return to Eqn.~\eqref{eq:33}, which may be written in terms of the nonlinear matrix $\mathcal{M}$ as $\sigma = \mathcal{P}^T\mathcal{M} \mathcal{G}\sigma_0 \mathcal{G}^T\mathcal{M}^T\mathcal{P}$ so that \begin{equation}
    \langle\sigma\rangle = \mathcal{P}^T\mathcal{M} \left[\int_{-\infty}^\infty f(\delta\gamma)(\mathcal{G}\sigma_0 \mathcal{G}^T)d(\delta\gamma)\right]\mathcal{M}^T\mathcal{P}
\end{equation}
since $\delta\gamma$ only appears in the $\mathcal{G}$ matrix. $(\mathcal{G}\sigma_0 \mathcal{G}^T)$ is a $n(m+1)\times n(m+1)$ matrix comprising an $(m+1)\times (m+1)$ block matrix of $n\times n$ sub-matrices that are each  $\sigma_0\delta\gamma^q$, where $q = i+j$, with $i$ the row and $j$ the column indices (starting at 0) of the $(m+1)\times(m+1)$ block matrix, i.e. for $m=3$, 
\begin{equation}
   (\mathcal{G}\sigma_0 \mathcal{G}^T) =\begin{bmatrix}
    \sigma_0 & \sigma_0\delta\gamma & \sigma_0\delta\gamma^2 & \sigma_0\delta\gamma^3\\
    \sigma_0\delta\gamma & \sigma_0\delta\gamma^2 & \sigma_0\delta\gamma^3 & \sigma_0\delta\gamma^4\\
    \sigma_0\delta\gamma^2 & \sigma_0\delta\gamma^3 & \sigma_0\delta\gamma^4 & \sigma_0\delta\gamma^5\\
    \sigma_0\delta\gamma^3 & \sigma_0\delta\gamma^4 & \sigma_0\delta\gamma^5 & \sigma_0\delta\gamma^6
    \end{bmatrix}\;.
\end{equation}

When integrated over $\delta\gamma$ for the normal distribution given in Eqn.~\eqref{eq:normal}, we can express the elements of the resulting block matrix, $\Sigma$ as 
\begin{equation}
    \Sigma_{ij} = \begin{cases}
    \langle\delta\gamma^2\rangle^{\frac{i+j}{2}}\sigma_0(i+j-1)!!& i+j\;{\rm even}\\
    0 &i+j\;{\rm odd}
\end{cases}\;,\label{eq:sigma}
\end{equation}
i.e. for $m=3$,
\begin{equation}
   \Sigma =\begin{bmatrix}
    \sigma_0 & 0 & \sigma_0\langle\delta\gamma^2\rangle & 0\\
    0 & \sigma_0\langle\delta\gamma^2\rangle & 0 & 3\sigma_0\langle\delta\gamma^2\rangle^2\\
    \sigma_0\langle\delta\gamma^2\rangle & 0 & 3\sigma_0\langle\delta\gamma^2\rangle^2 & 0\\
    0 & 3\sigma_0\langle\delta\gamma^2\rangle^2 & 0 & 15\sigma_0\langle\delta\gamma^2\rangle^3
    \end{bmatrix}\;.
\end{equation}
The new beam normalized emittance, defined as $\epsilon_N = \sqrt{\det{\sigma}}$  is, therefore,

\begin{equation}
    \epsilon_N = \sqrt{\det\left(\mathcal{P}^T\mathcal{M}{\Sigma}\mathcal{M}^T\mathcal{P}\right)}\;.\label{emittance_growth}
\end{equation}

\begin{figure}
    \centering
    \includegraphics[width=0.45\textwidth]{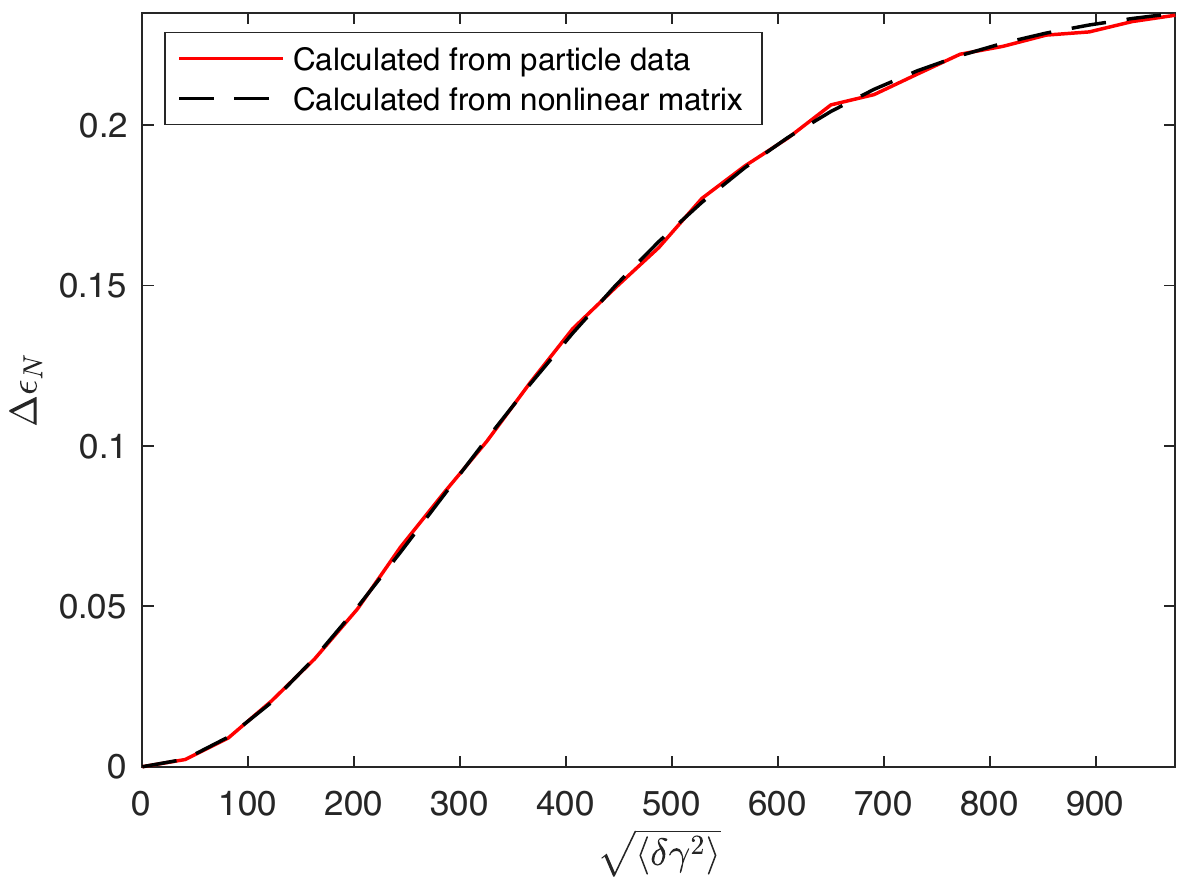}
    \caption{Normalized emittance growth through a nonlinear matrix calculated {by} summing over a large number of particle tracks (blue) compared the calculation using Eqn.~\eqref{emittance_growth} (red), as a function of $\sqrt{\langle\delta\gamma^2\rangle}$.
    }
    \label{fig:chromtest}
\end{figure}

To verify this expression, Fig.~\ref{fig:chromtest} shows the change in normalized emittance (i.e. subtracting the initial emittance) for a beam of particles with a mean energy $\gamma_0 = 19500$  for a range of values in the gaussian width of the energy distribution, $\sqrt{\langle\delta\gamma^2\rangle}$. The red curve shows the calculation of Eqn.~\eqref{emittance_growth} using the nonlinear transfer matrix expanded to $m=9$ orders (the matrix is that of first accelerating stage calculated in the next section) $\mathcal{M}$. This agrees with the data indicated by the blue curve, which shows the normalized emittance calculated using Eqn.~\eqref{emit:simps} for 100,000 individual particles with  energy offsets $\delta\gamma$ randomly sampled from a normal distribution. The small fluctuations in the blue curve are due to particle statistics.

{Eqn.~\eqref{emittance_growth} represents the growth in transverse emittance of a beam slice of width $\Delta\xi$ at a particular wake phase $\xi_j$. The chromatic emittance growth of a whole beam with longitudinal density profile $b(\xi)$, discretized as $b_j$ will be
\begin{equation}
    \epsilon_N = \sqrt{\det\left[\mathcal{P}^T\left(\sum_jw_j\mathcal{M}_j{\Sigma}_j\mathcal{M}_j^T\right)\mathcal{P}\right]}\;.\label{emittance_growth_beam}
\end{equation}
where the weights $w_j$ are the terms in the discrete integral of the beam profile, e.g. for Riemann summation $w_j = b_j\Delta\xi$.
}
\section{\label{sec:level4} Other transport elements}
The main advantage of using a  transfer matrix approach for the plasma elements is to be able to combine it with other elements. For a drift space of length $L$, the vector $w$ is transformed by the matrix $$d=\exp(A_1L/c)=\mathbb{I}_n+A_1 \frac{L}{c}\;.$$

For the nonlinear vector $w_\delta$, the corresponding drift space matrix can be found using the same process as before, by expanding $d$ in a Taylor series up to highest term $m$ in $\delta\gamma$ around design energy $\gamma$. As $d$ only contains terms in $1/\gamma$ and constant with respect to $\gamma$, as with the plasma accelerator transfer matrix, the nonlinear transfer matrix for a drift space is
\begin{equation}
\mathcal{D} = \mathbb{I}_{m+1}\otimes d + {\Gamma}\otimes A_1 \frac{L}{c}\;.
\end{equation}

For introducing focusing optics to the system, in the context of plasma accelerators these may be conventional optics, i.e. quadrupole triplets, or plasma optics. Given a transfer matrix through the optic, a nonlinear matrix up to order $m$ can always be derived through the process outlined previously. For simplicity, here we consider a general optic using the thin lens approximation,
{which is valid} provided the effective focal length is very large compared to the beam size. Whether the optic is a quadrupole triplet or some sort of plasma lens, however, it will be chromatic. We assume that the lens focal length $f$ has a $f\propto\gamma$ relationship. Therefore, given the transfer matrix
\begin{equation}
    F = {\mathbb{I}_{n/2}\, \otimes }
    \begin{bmatrix}
    1 &0 \\ -\frac{c}{f^\star}&1
    \end{bmatrix}
\end{equation}
where $f^\star = f(\gamma)/\gamma$ is the focal length divided by $\gamma$ and is therefore a constant, the nonlinear focusing matrix is trivial, as {it} has no {explicit} $\gamma$ dependence,
\begin{equation}
    \mathcal{F} = \mathbb{I}_{m+1}\otimes  F\;.
\end{equation}
{Note that using the coordinates we choose here, $(x,u_x)$ instead of $(x,x^\prime)$, the chromatic effects of focusing manifest themselves in the drift-space matrix rather than the lens matrix.}
\section{\label{sec:10gev}Particle-in-cell simulation of a 10 GeV stage}
For demonstrating the nonlinear transfer matrix approach outlined in this paper, we use the 3D relativistic particle-in-cell framework OSIRIS 4.0 \cite{osiris_paper} to simulate a beam driven plasma wakefield accelerator.  A beam driven plasma wakefield was chosen for clarity in this paper, but this technique would be more interesting for a laser driven wakefield, in which the wake evolves as the laser pulse propagates due to self-phase modulation etc. There is no limitation on the complexity of the laser evolution, as the particle beam is at fixed phase relative to the moving box, so the laser may fall back due to dispersion or modulate in a complex way, but the fields at the particle position will be accurately captured (provided the particle beam is ultrarelatistic, $\gamma>1000$ from the beginning, which means that an injection stage of a plasma accelerator cannot be  modelled accurately using this technique).

\subsection{Simulation description}
A simulation was run on an $z\times x\times y$ mesh of $128\times 150\times 150$ grid points with spatial limits from $-10c/\omega_{p0}$ to $10c/\omega_{p0}$ in the transverse dimensions and $-14c/\omega_{p0}$ to $2c/\omega_{p0}$ in the $z$ direction with a time-step $\omega_{p0}\Delta t = 0.06$. Standard 5-pass smoothing algorithms were applied to the electromagnetic fields and currents. A 2nd order dual type  electromagnetic solver \cite{Li_CPC_2021}  and open (perfectly matched layer and open particle bounds) boundary conditions were used. Two species were included; a driver beam species of electrons with 4 particles-per-cell and $\gamma = 80,000$ with a gaussian shape in all directions having a peak density of 150 $n_0$ and widths $\sigma_{z} = 0.7c/\omega_{p0}$ and $\sigma_{x},  \sigma_{y} = 0.2c/\omega_{p0}$, and a plasma species of electrons with 4 particles-per-cell, a peak density of $n_0$ and a profile given by the function $n(z) = \exp(-[(z-8000c/\omega_{p0})^2/(7000c/\omega_{p0})^2]^{10})$. For a density $n_0 = 10^{16}\;{\rm cm}^{-3}$, this corresponds to an 80 cm long plasma channel with approximately 8 cm long ramps in density from vacuum. Long ramps have been shown to help with adiabatic matching of the beam \cite{Xu_PRL_2016}. The drive beam was started in vacuum with zero charge and momentum, and was both accelerated in $z$ and ramped up in charge at the start of the simulation, with the equations of motion otherwise fixed, to establish the correct vacuum fields before entering the plasma.

\begin{figure}[htbp]
\begin{center}
\includegraphics[width =0.5\textwidth]{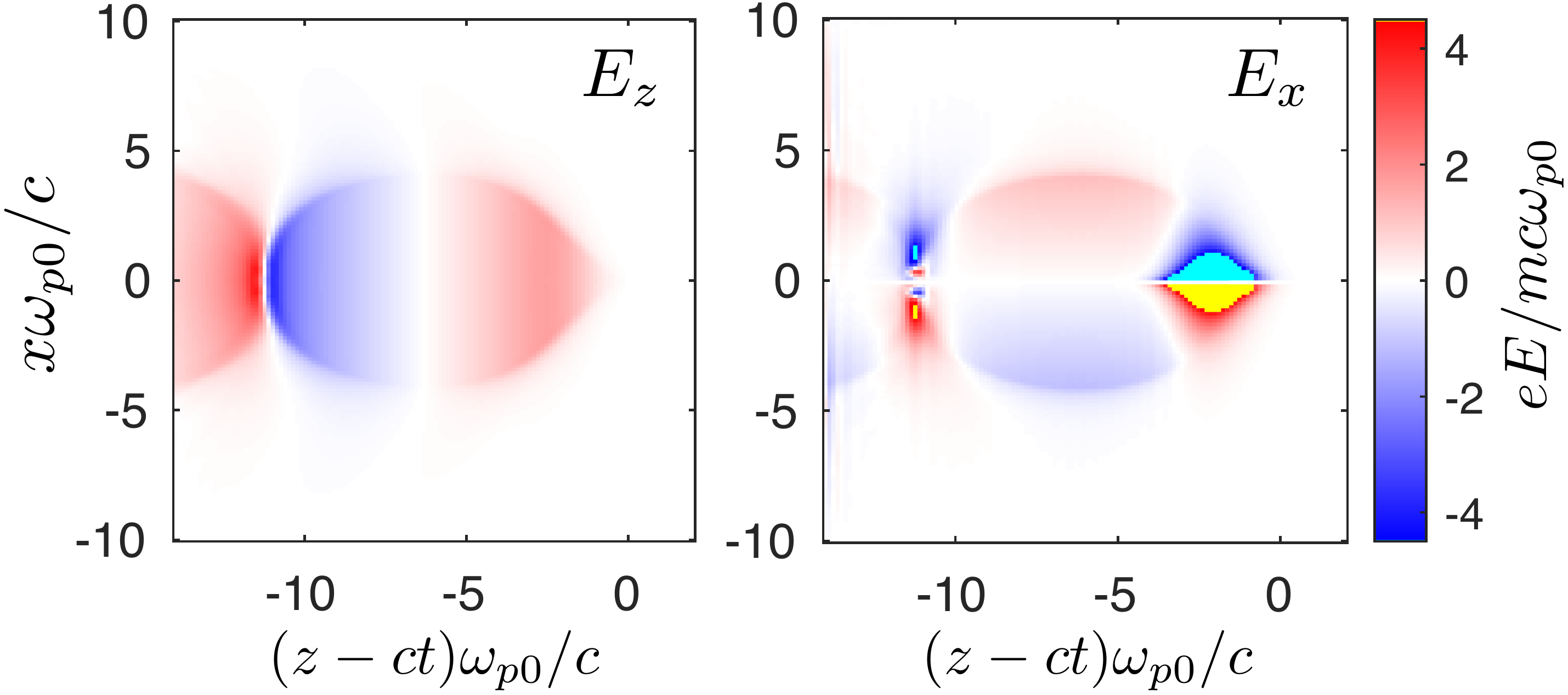}
\caption{Fields taken from the 3D simulation at $\omega_{p0}t=9000$ in the $z$-$x$ plane at $y=0$. (left) Accelerating ($E_z$) field and (right)  focusing ($E_x$) field.  }
\label{fig:fields_bubble}
\end{center}
\end{figure}

Fig.~\ref{fig:fields_bubble} shows the electric fields taken from the 3D simulation at $\omega_{p0}t=9000$ in the $z$-$x$ plane at $y=0$. (left) the accelerating ($E_z$) field and (right) the focusing ($E_x$) field. The cyan and yellow colors are because the colormap is saturated, due to the strong fields where the drive beam is located.

\begin{figure}[htbp]
\begin{center}
\includegraphics[width =0.5\textwidth]{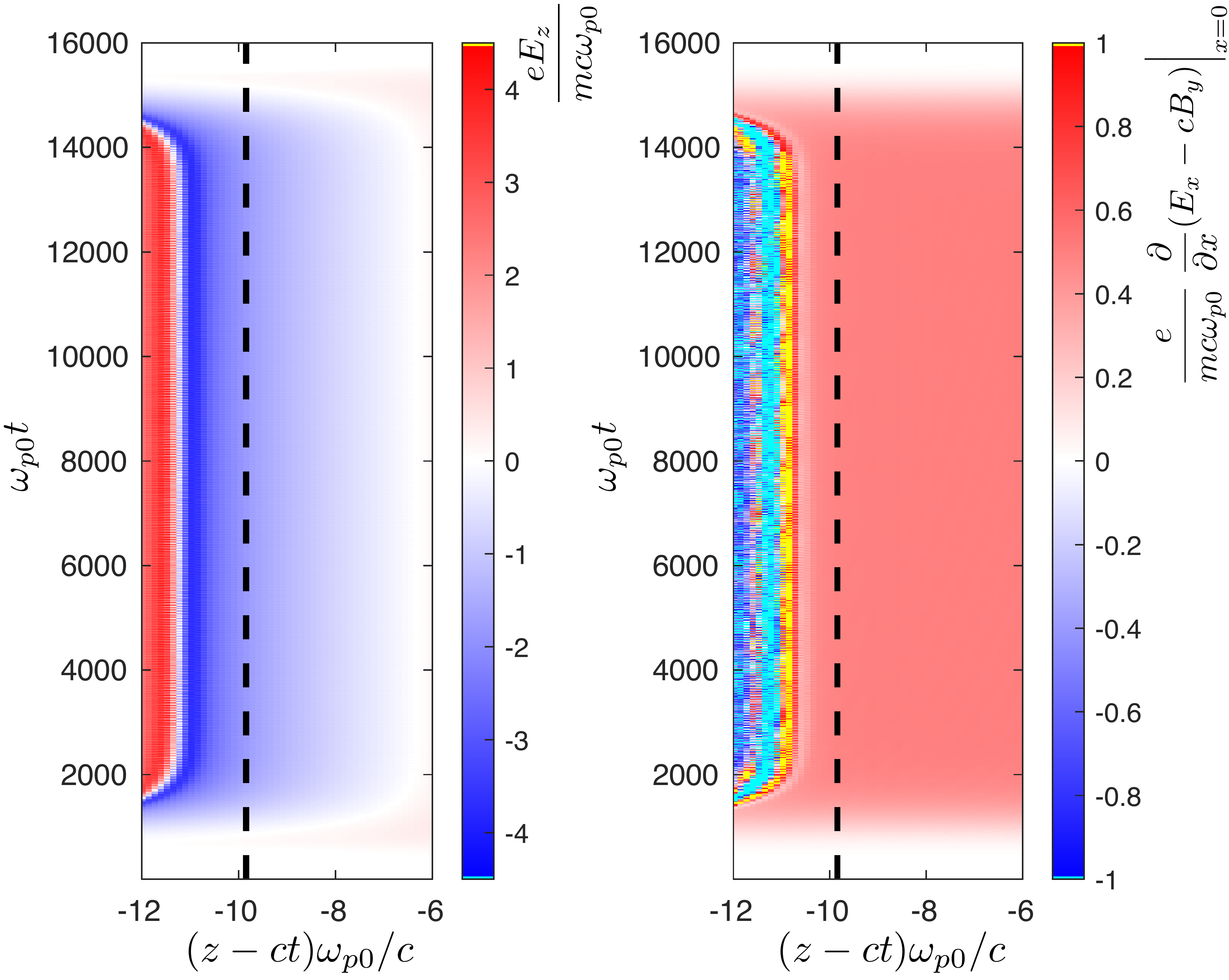}
\caption{(left) The axial accelerating field $E_z$ and (right)  focusing gradient on an ultrarelativistic particle propagating in the $z$ direction, $(e/mc\omega_{p0})\partial(E_x - cB_y)/\partial x|_{x=0}$ both extracted as a lineout in the $z$ direction along the axis and plotted as a time series. The black dashed line indicates the wake phase chosen for the beam transport.}
\label{fig:fields_of_time}
\end{center}
\end{figure}

Line diagnostics extracted the $E_z$, $E_x$, and $B_y$ fields along $z$ direction on the mesh points either  side of the beam center to obtain the field gradients. These were extracted every 10 timesteps, i.e. $0.6/\omega_{p0}$. The $E_x$ and $B_y$ fields were used to calculate the field gradient along the axis in the $x$ direction by subtracting the values either side of the center line and dividing by $2\Delta x$, i.e. the center differenced finite difference gradient
\begin{eqnarray}
    \left[\alpha_{xx}^2\right]^n_j &=& -\left.\frac{\partial}{\partial x}\left(E_x - cB_y\right)\right|_{x=0,j}^n \\\nonumber
    &\simeq& -\frac{(E_x - cB_y)^n_{j,k+1,l} - (E_x - cB_y)^n_{j,k-1,l}}{2\Delta x}\;,
\end{eqnarray}

where the indices $j,k,l$ are the grid indices expressed relative to the center line and $n$ is the time step. The accelerating field $E_z$ and focusing force gradient as a function of time throughout the whole simulation are shown in Fig.~\ref{fig:fields_of_time}. 
These can be used to generate nonlinear matrix $\mathcal{M}$ for propagation of a beam with initial beam energy $\gamma_0 mc^2$ through the whole simulation by using the methods described in the previous section. In principle this could be performed for every wake phase. Here, we choose only the wake phase indicated by the black dashed line in Fig.~\ref{fig:fields_of_time}.

\subsection{Transport through 10 GeV stages}
\begin{figure}[htbp]
\begin{center}
\includegraphics[width =0.5\textwidth]{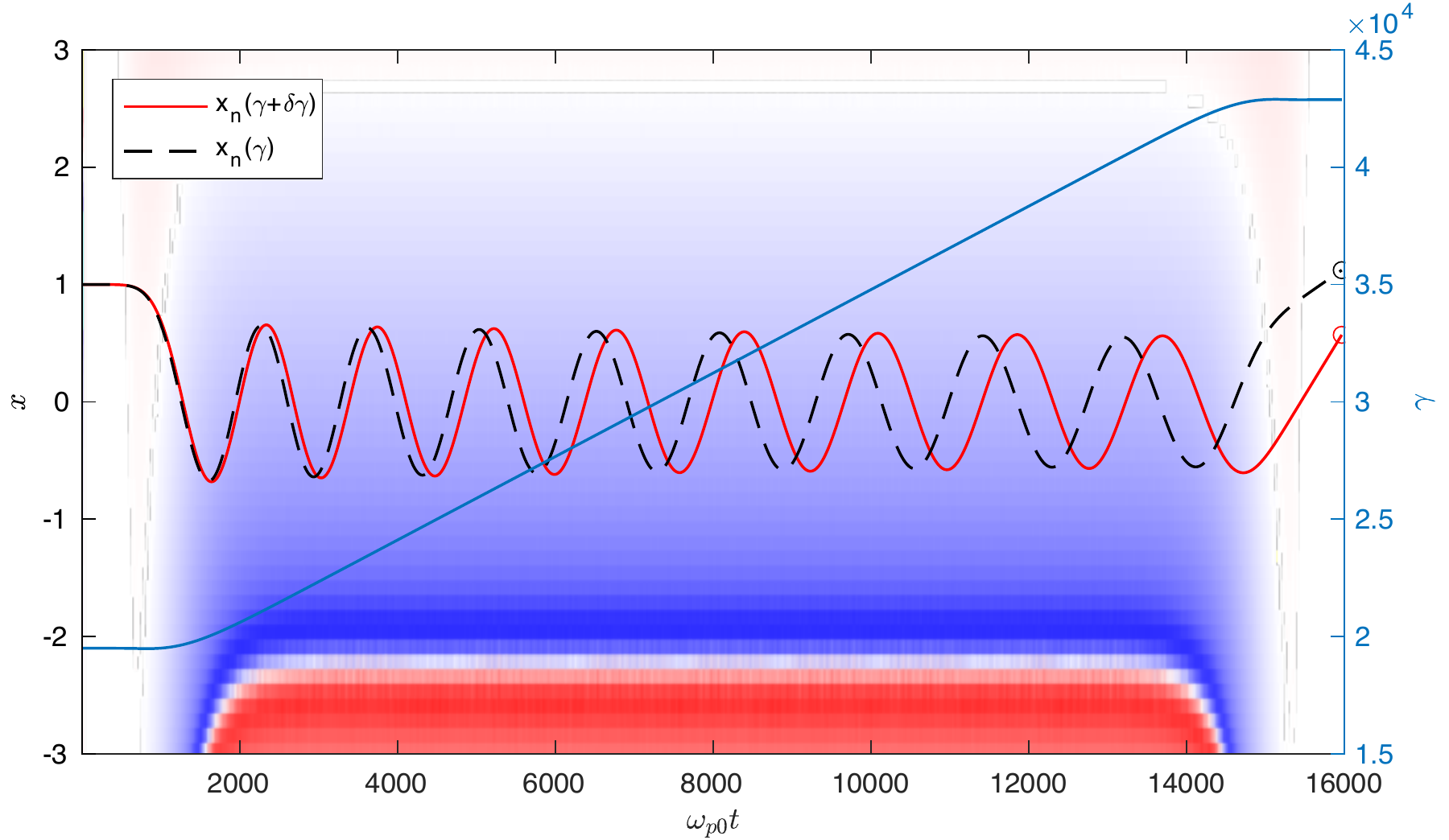}
\caption{Representative particle track through fields taken from the 3D simulation. The design (normalized) energy is initially $\gamma = 19500$, and it is accelerated to $\gamma = 42891$, as indicated by the blue curve. The particle undergoes betatron oscillations indicated by the red and black dashed curves. The red curve shows the track for a beam $\delta\gamma$ from the design energy, with  $\delta\gamma = 1950$. The black and red circled dots at the end show the result of a calculation from the initial coordinates using the combined matrix $\mathcal{M}$. Colorscale is the accelerating field (see Fig.~\ref{fig:fields_of_time}). }
\label{fig:bets}
\end{center}
\end{figure}

Using the transverse field gradient and longitudinal field time histories obtained from the particle-in-cell simulation, we can construct the matrix $\mathcal{M}^n(\gamma^n)$, at every time step and then calculate the composite matrix $\mathcal{M}(\gamma_0)$ comprising transport through an entire plasma accelerating stage for a given initial energy $\gamma_0$. Fig.~\ref{fig:bets} shows a representative trajectory through the fields. The red and black lines show repeated application of  $\mathcal{M}^n(\gamma^n)$ to initial extended coordinates $w_{\delta 0}$, for either $\delta\gamma = 0$ or $\delta\gamma = 1950$ for an initial beam energy $\gamma=19500$, i.e. $\delta\gamma/\gamma = 0.1$. The blue curve shows the particle energy as a function of propagation time, up to $\gamma = 42891$, i.e. just over 10 GeV acceleration in the stage. Although it is obvious has to be the case, we also explicitly show the result of $\mathcal{M}(\gamma_0)w_{\delta0}$ as a red and black circle, showing  the transformation of the particle coordinates using the single $\mathcal{M}(\gamma_0)$. 

We may now proceed to designing an accelerating lattice by using the final energy $\gamma_i$ for each stage and using it as the initial energy for the next stage. Through this iterative process, we end up with a set of nonlinear matrices $\mathcal{M}^s$, where $s$ is the stage index, i.e. stage 0 is $\mathcal{M}^0 = \mathcal{M}(\gamma_0)$ etc. This matrix set can then be used to integrate with other transport elements.

\section{Design of a simple 1 TeV lattice}\label{sec:focusing}
To illustrate the use of these plasma accelerator transfer matrices, we introduce a simple lattice design, as illustrated in Fig.~\ref{fig:Lattice}. Each accelerator stage has a focusing optic after it. A thin lens is assumed for this focusing optic{, with focal length at the $s$th stage given by $f_{Ls}=1/{(k_sL_s)}$ where $L_s$ is the lens thickness and $k_s$ represents the lens strength. Since for the thin lens to be valid, $L_s \ll f_{Ls}$, we must have  $L_s = \epsilon f_{Ls}$, where $\epsilon$ is a small number. Therefore, we scale the focal length using $f_{Ls} = 1/{(k_s\epsilon f_{Ls})}$. Assuming a fixed field gradient, the lens strength is inversely proportional to $\gamma$, $k_s\propto 1/\gamma$, and so the focal length should scale as $f_{Ls}\propto \sqrt{\gamma}$ to maintain the thin lens approximation for all stages. We start with a first stage focal length at the design energy of $f_{L1} = 8000c/\omega_{p0}$, corresponding to approximately 40 cm for a plasma density of $10^{16}$~cm$^{-3}$. The focal lengths of the optics in subsequent stages scale as $f_{Ls} = f_{L1}\sqrt{\gamma_{fs}/\gamma_{f1}}$, where $\gamma_{fs}$ is the energy after the $s$th stage.}

Each plasma accelerating stage has its own focusing characteristics, and may act as either a positive or negative lens, depending on the betatron phase. It would be possible to tune the betatron phase through each acceleration by adjusting the plasma length, but here we simply use an adjustable drift distance between lenses / accelerating stages to have ``$2f$'' re-imaging of the beam to each stage. The focal length of the thin lenses is sufficiently long that the distance between the lens and accelerating stage remains positive even if the accelerating stage acts as a negative lens (i.e. having a negative drift to the virtual focus).

The distances between each accelerating stage and lens are calculated in the following way.
\begin{enumerate}
    \item The accelerating stage acts like a thick lens, so the {distances to the} primary (FPP) and secondary (SPP) principal planes are calculated through 
    $$
    d_{FPP_s} = \frac{1 - \mathcal{M}_{2,2}^s}{\mathcal{M}_{2,1}^s}\gamma_{s-1}
    $$
    and
    $$
    d_{SPP_s} = \frac{1 - \mathcal{M}_{1,1}^s}{\mathcal{M}_{2,1}^s}\gamma_{s}\;,
    $$
    where $\gamma_{s}$ is the \emph{final} beam energy after acceleration through the $s$th stage.
    \item This allows correction of $\mathcal{M}^s$ to act as a thin lens through 
    $$
    \mathcal{M}^{\star s} = \mathcal{D}_{SPP_s}\mathcal{M}^{ s}\mathcal{D}_{FPP_s}\;,
    $$
    where $\mathcal{D}_{FPP_s}$ is the nonlinear matrix for the drift space for distance $d_{FPP_s}$ and similar for $SPP$.
    \item The effective focal length of the accelerating stage is $f_s = -\gamma^s/\mathcal{M}_{2,1}^s$. We express the  drift space with length $2f_s$ as $\mathcal{D}_{2f_s}$. 
    \item The thin lens focusing optic with focal length $f_L$ has a matrix $\mathcal{F}_L^s$, and we express the  drift space with length $2f_L$ as $\mathcal{D}_{2f_L}^s$. These need an $s$ index because they depend on the beam energy $\gamma^s$. 
    \item The matrix describing a ``cell'' of the lattice, comprising the $s$th accelerating stage and focusing optic separated by ``$2f$'' distances is therefore
    $$
    \mathcal{C}^s = \mathcal{D}_{2f_L}^s\mathcal{F}_L^s\mathcal{D}_{2f_L}^s\mathcal{D}_{2f_s}\mathcal{M}^{\star s}\mathcal{D}_{2f_{s-1}}\;.
    $$
\end{enumerate}
The $\mathcal{C}^s$ matrices can then be combined to form 
\begin{equation}
    \mathcal{C} = \prod_{s=0}^{N_{\rm stages}} \mathcal{C}^s\;,
\end{equation}
which is the matrix that describes transport through the full accelerating structure.
\section{Chromatic emittance growth through 85 stage plasma accelerating lattice}
The matrix $\mathcal{C}$ was calculated for the plasma accelerating stage simulated in Section \ref{sec:10gev} for 85 stages / focusing lenses in the arrangement described in Section \ref{sec:focusing}, each accelerating the beam by 11.95 GeV to a maximum energy of 1.03 TeV. The particle distribution is initialized  using a beam matrix $\sigma_0 = \begin{bmatrix}{\epsilon_{N0}}^2&0\\0&1\end{bmatrix}$, which corresponds to an input  beam with $\theta\sim1/\gamma$ convergence angle  focused at $2f_0$ before the start of the first accelerating section. 
\begin{figure}[htbp]
\begin{center}
\includegraphics[width =0.5\textwidth]{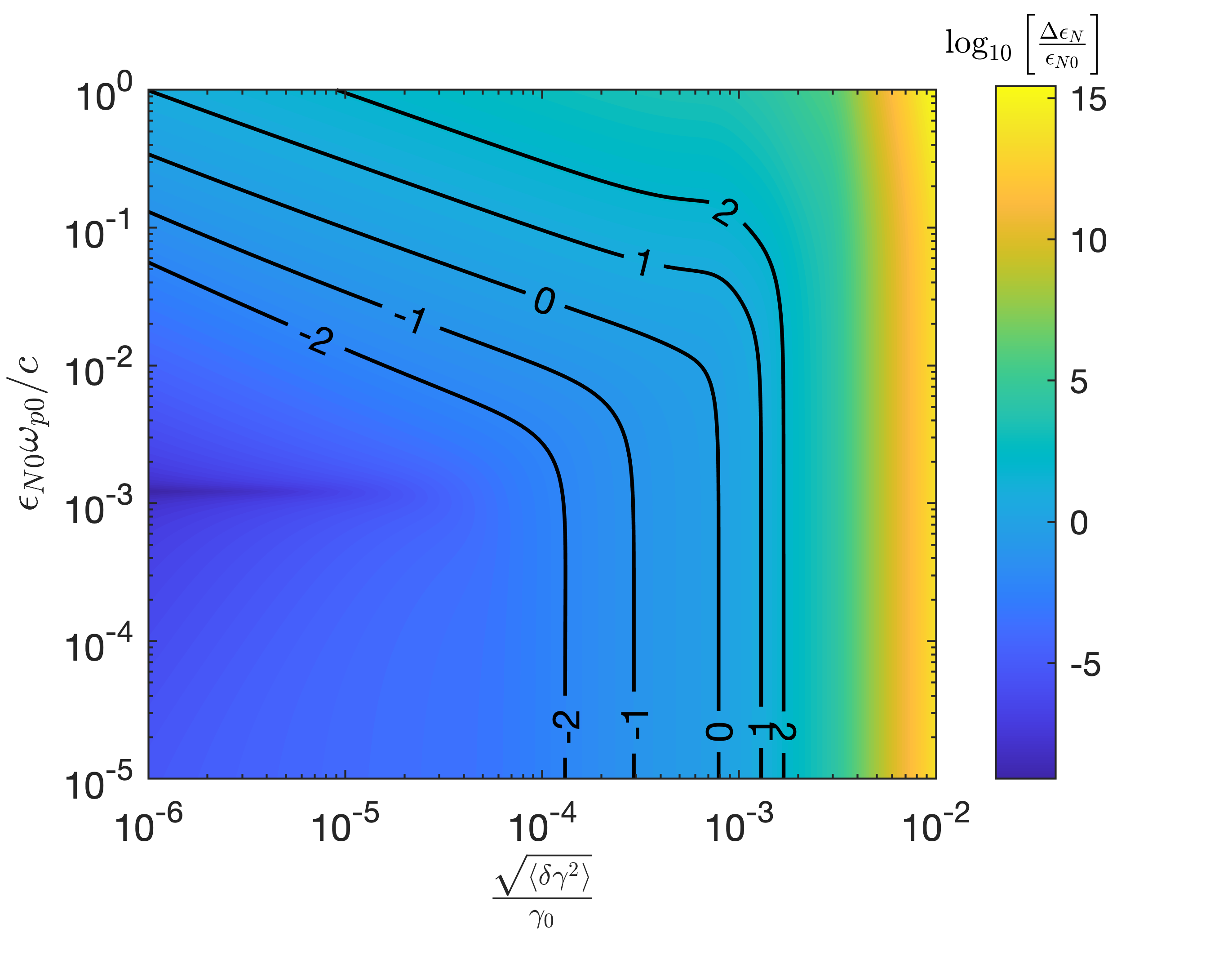}
\caption{Relative emittance growth $\Delta\epsilon_{N}/\epsilon_{N0}$ through 85 accelerating stages as a function of {\emph{initial}} relative energy spread $\sqrt{\langle\delta\gamma^2\rangle}/\gamma_0$ and initial normalized emittance $\epsilon_{N0}$. The colormap / contours show the base-10 logarithm of  $\Delta\epsilon_{N}/\epsilon_{N0}$. }
\label{fig:emitt_grow_final}
\end{center}
\end{figure}

{
\subsection{Chromatic slice-emittance growth}}
Fig.~\ref{fig:emitt_grow_final} shows the relative {(slice)} emittance growth $\Delta\epsilon_{N}/\epsilon_{N0}$ through this lattice {at the phase indicated in Fig.~\ref{fig:fields_of_time}} as a function of {\emph{initial}} relative energy spread $\sqrt{\langle\delta\gamma^2\rangle}/\gamma_0$ and initial normalized emittance $\epsilon_{N0}$, calculated using Eqn.~\eqref{emittance_growth}. The colormap / contours show the base-10 logarithm of  $\epsilon_{N}/\epsilon_{N0}-1 =\Delta\epsilon_{N}/\epsilon_{N0}$. The normalized emittance $\epsilon_N$ is normalized to the length unit $c/\omega_{p0}$, which means that for a baseline plasma density of $10^{16}$~cm$^{-3}$, a normalized emittance of $\epsilon_N\sim$mm-mrad ($\mu$m) corresponds to $\epsilon_{N}\omega_{p0}/c = 0.019$. This parameter search indicates that for initial relative energy spread below $\sqrt{\langle\delta\gamma^2\rangle}/\gamma_0\lesssim 10^{-3}$ and  initial normalized emittance below $\epsilon_{N}\omega_{p0}/c\lesssim10^{-2}$ (i.e. $\epsilon_N\lesssim$ mm-mrad), the  chromatic emittance growth is relatively small ($\Delta\epsilon_{N}/\epsilon_{N0}\lesssim \epsilon_{N0}$).
{
\subsection{Energy spread growth of a finite length beam}
The example before calculated the evolution of the transverse phase space for an ensemble of particles at a particular wake phase, which therefore experiences \emph{no energy-spread growth} as the particles all interact with an identical longitudinal electric field. The effect of a finite duration beam, which experiences different accelerating fields at different phases in general, can be taken into account by calculating the transport for different phases $\xi$ and combining the resulting beam matrices, as in Eqn.~\eqref{emittance_growth_beam}. One very important consideration is loading of the wake \cite{Katsouleas_PA_1987,Tzoufras_PRL_2008}, which requires a specially shaped bunch. The effect is to flatten the electric field experienced by the witness bunch such that in the ideal case all particles experience the same acceleration and therefore no energy-spread growth as in the previous example. This effect can be included  using our method by the addition of an ultrarelativistic witness beam in the simulation, as in the ultrarelativistic limit its fields do not depend on the beam energy and cancel for co-propagating particles of the same charge. 

In practice, however, perfect loading will not be possible and, in general, some energy-spread growth will be expected. We leave detailed studies of beam loading, beam shape and duration using calculations of the beam matrix over a range of phases for future work, but it is instructive to redo the calculation from the previous section to include energy-spread growth effects and estimate how good the beam-loading must be. We do this using the following simple model:

Assuming a finite-duration beam, with width $\sqrt{\langle\delta\xi^2\rangle}$ in $\xi$, that has initially zero \emph{slice} energy-spread (in practice, it just needs to be much smaller than the overall beam energy spread) and gains energy spread at each step because particles at different phases experience different accelerating fields. The normalized charge-density profile of the beam is $b(\delta\xi)$, where $\delta\xi = \xi - \xi_0$ and $\xi_0$ is the phase of the beam having reference energy $\gamma_0$. We assume that the energy of the beam at (relative) phase $\delta\xi$ is given by an arbitrary function $g(\delta\xi)$. Hence, the $\delta\gamma$-$\delta\xi$ phase-space distribution of the beam is described by the distribution
\begin{equation}
f(\delta\gamma,\delta\xi) = b(\delta\xi)\delta(\delta\gamma - g(\delta\xi))\;,
\end{equation}
where $\delta(x)$ is the Dirac delta distribution. The energy distribution of the full beam is $\rho(\delta\gamma) = \int_{-\infty}^\infty f d(\delta\xi)$, which can be written as 
\begin{equation}
\rho(\delta\gamma) = \int_{-\infty}^\infty \frac{b(\delta\xi)}{|g^\prime(g^{-1})|}\delta(\delta\xi - g^{-1})  d(\delta\xi)\;,
\end{equation}
where $g^{-1}(\delta\gamma)$ is the inverse of the function $g(\delta\xi)$, i.e. $g^{-1}(\delta\gamma) = \delta\xi$ and the prime $\prime$ indicates the derivative with respect to $\delta\xi$.  Hence, 
\begin{equation}
\rho(\delta\gamma) = \frac{b(g^{-1}(\delta\gamma))}{|g^\prime(g^{-1}(\delta\gamma))|}\;.
\end{equation}
In the blowout regime, the transverse fields are uniform in $\xi$ and the longitudinal  field is linear in the interior. We therefore assume that over a timestep, the function is given by $g(\delta\xi) = \alpha \delta\xi$, where $\alpha$ is a constant, i.e. a linear chirp, and the bunch shape is gaussian, i.e. 
\begin{equation}
b(\delta\xi) =  B \exp\left(-\frac{\delta\xi^2}{2\langle\delta\xi^2\rangle}\right)\;,
\end{equation}
where $B$ is a normalizing constant and $\sqrt{\langle\delta\xi^2\rangle}$ is the bunch longitudinal width.  The energy distribution of the whole beam is therefore
\begin{equation}
\rho(\delta\gamma) = \frac{B }{|\alpha|}\exp\left(-\frac{\delta\gamma^2}{2\alpha^2\langle\delta\xi^2\rangle}\right)\;.
\end{equation}
This distribution is identical to that used in Eqn.~\eqref{eq:normal}, with an energy spread $\sqrt{\langle\delta\gamma^2\rangle} = \alpha\sqrt{\langle\delta\xi^2\rangle}$  and therefore the same expanded beam matrix can be used, under the assumption that the transverse fields do not vary over the beam (i.e. either it is ultrashort or in the fully blown out regime. The important difference is, however, that the beam energy-spread changes every timestep, because of the variation in accelerating field throughout the beam.

Under the assumption that the field is linear, the increase (or decrease \cite{Pathak_POP_2021}) in the beam energy spread at each time-step will be because of a change in the beam chirp, i.e.
\begin{equation}
\sqrt{\langle\delta\gamma^2\rangle}(t) = q\sqrt{\langle\delta\xi^2\rangle}\int_0^t\left.\frac{\partial E_z}{\partial \xi}\right|_{\xi = \xi_0}(t^\prime)dt^\prime\;, 
\end{equation}
The gradient in $E_z$ does not in general have to follow the same temporal evolution as $E_z$. Here, for simplicity, we assume that the gradient of the longitudinal field evolves identically to the field, i.e. $ \partial E_z/\partial \xi\propto E_z$, which implies that the ratio of the beam energy-spread to the beam energy, $\sqrt{\langle\delta\gamma^2\rangle}/\gamma_0$, is a constant.

\begin{figure}[htbp]
\begin{center}
\includegraphics[width =0.5\textwidth]{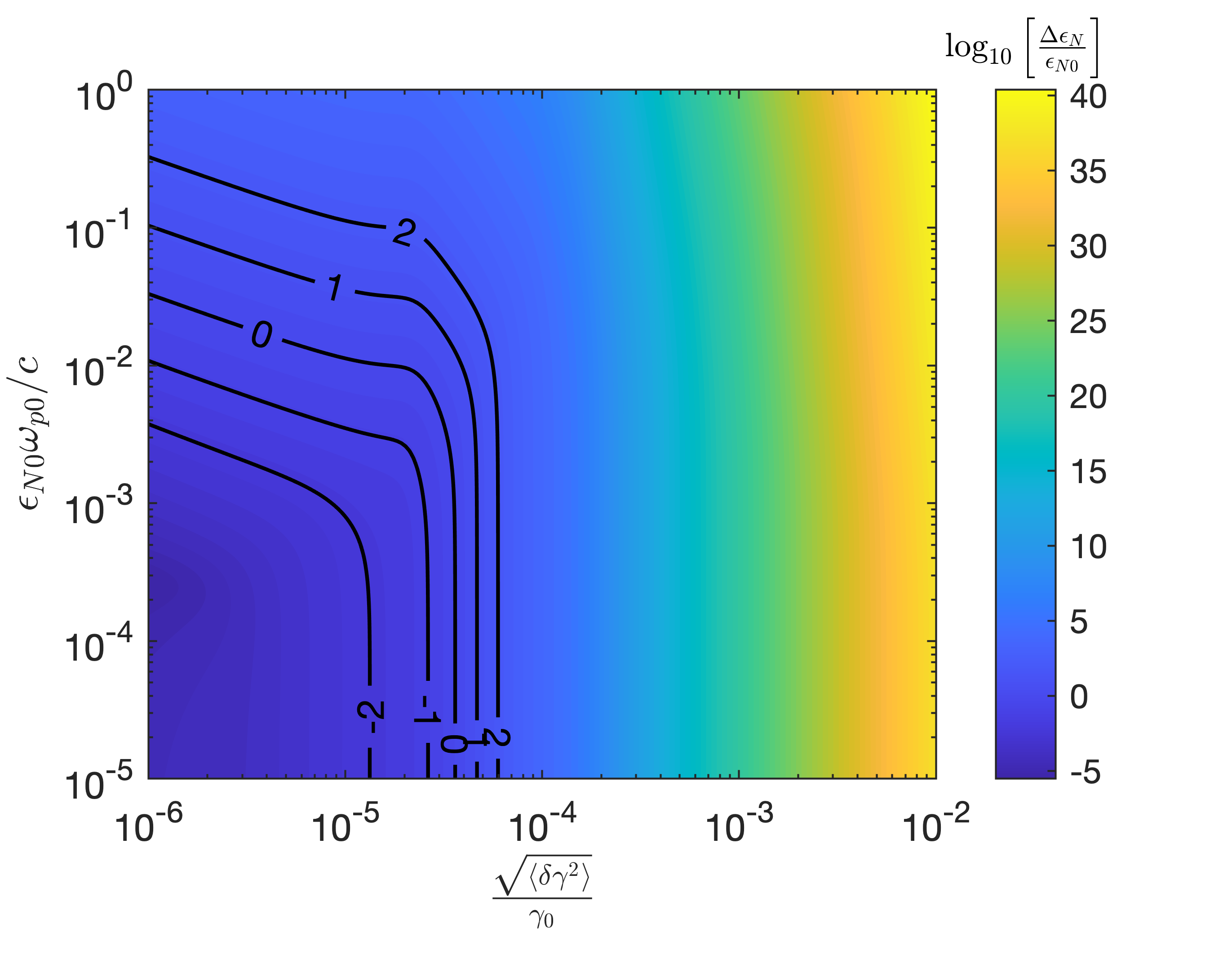}
\caption{{Relative emittance growth $\Delta\epsilon_{N}/\epsilon_{N0}$ through 85 accelerating stages as a function of \emph{constant} relative energy spread $\sqrt{\langle\delta\gamma^2\rangle}/\gamma_0$ and initial normalized emittance $\epsilon_{N0}$. The colormap / contours show the base-10 logarithm of  $\Delta\epsilon_{N}/\epsilon_{N0}$. }
}
\label{fig:emitt_grow_dgog}
\end{center}
\end{figure}

By removing the $\gamma$ factors in $\Gamma$, Eqn.~\eqref{gammafactor}, and replacing the $\delta\gamma$ factors in $\Sigma$, Eqn.~\eqref{eq:sigma}, and the extended system vector $w_\delta$, Eqn.~\eqref{eq:wdelta}, etc. with $\delta\gamma / \gamma$, the transfer matrix approach with an expansion in (constant) $\delta\gamma/\gamma$ instead of (constant) $\delta\gamma$ can be used to calculate the chromatic emittance growth of a finite duration gaussian beam on a field gradient that results in energy spread growth through the development of a linear chirp in the beam. For the same 85 stage lattice used in the previous example, the parameter space of the emittance growth is shown in Fig.~\ref{fig:emitt_grow_dgog}. These data indicate that for a growth in the energy spread owing to beam chirp below $10^{-5}$ of the energy gain per stage, i.e. $\sqrt{\langle\delta\gamma^2\rangle}\lesssim 10^{-5}\gamma_0 $ and  initial normalized emittance below $\epsilon_{N}\omega_{p0}/c\lesssim10^{-2}$, i.e. $\epsilon_N\lesssim$ mm-mrad, the  chromatic emittance growth is relatively small ($\Delta\epsilon_{N}/\epsilon_{N0}\lesssim \epsilon_{N0}$). This requires either a short beam or flattened fields through beam loading such that the variation in the accelerating field over the bunch is of order $10^{-5}$ of the field strength.
}
\section{Conclusions}
In this work we have calculated the transport of an electron beam through an 85 stage beam driven plasma accelerator using simple lenses for beam transport. No effort was made to optimize this (by for example paying attention to the betatron phase at the exit of each stage by tuning the plasma accelerator length), but nevertheless {negligible} chromatic emittance growth {can be achieved}  for likely collider parameters even for this simple design. {In particular, we have shown that limiting the energy spread growth due to beam chirp from non-uniform fields is an important consideration, and beamloading to flatten the fields alone may be challenging. The use of a plasma dechirper \cite{DArcy_PRL_2019,Pathak_POP_2021} is one way to meet this challenge, by correcting for energy spread growth as the bunch is accelerated.} As has been previously shown, the effects of beam misalignment need consideration \cite{Thevenet_PRAB_2019}, which could be studied through a realistic design using the  linear transfer matrix approach outlined here and a large number of particles. The choice not to model e.g. real quadrupoles was because it is not clear what the optics will be for a plasma collider design. If they are plasma optics, then these could be simulated and modeled using the approach described here with the nonlinear matrices. 

Note that a beam driven accelerator {was chosen here, as it is a clearer example for demonstration purposes}, but the real usefulness of this technique will be for laser-pulse driven wakes where the laser can have a complex interaction with the plasma leading to a relatively highly dynamically evolving wakefield. This is no problem for the transfer matrices here as the full evolution will be captured. Indeed, for lepton beams sufficiently energetic that their phase slippage would be small compared to the wavelength of the drive beam  --- e.g. for 1 $\mu$m lasers, over a meter length stage this is already a reasonable approximation for only a 1 GeV electron beam --- then even the interaction of the particle with the {oscillating laser fields themselves} would be correctly modeled.

It is worth  briefly reviewing the approximations of this transfer matrix approach compared with a full scale simulation to understand its limitations. The main approximations can be summarized as follows:
\begin{itemize}
    \item \textit{The constant speed of light phase approximation.} This is generally a good approximation. For particles with energies exceeding a GeV, the phase advance owing to this approximation over a meter propagation (a typical plasma stage length) is a fraction of a tenth of a micron, and at 10 GeV it is already at the nanometer range. This means that even interaction with the fast oscillating laser fields should be accurate in most cases.
    \item \textit{The paraxial approximation.} This is also generally a good approximation for scenarios of interest. This means that the beam is primarily traveling in the forward direction and also that the transverse forces are linear. While the fields in a plasma accelerator can be nonlinear in general, they will be linear near the axis. The beam will be required to be small compared to the wake diameter for any reasonable staged design to work. If non-linearities were necessary to consider, they could be included as a perturbation using a nonlinear matrix approach similar to the one we address earlier in this paper for energy spread (i.e. by including terms in $x^2,y^2$ etc. in the matrix).
    \item \textit{The external fields approximation.} This is a limitation of the model in addressing problems such as instabilities like hosing \cite{Mehrling_PRL_2017}, since the particle's currents do not feed-back onto the wake fields. Another prominent issue to address is that of beam loading \cite{Tzoufras_PRL_2008}, i.e. that the particle bunch in a plasma wakefield may be designed to flatten the longitudinal field so that particles at different phases experience the same accelerating gradient. However, this approximation does avoid fictitious numerical feedback between the beam and the fields \cite{Lehe_PRZ_2013}.

\end{itemize}
    For properly modeling instabilities involving the interaction of the beam with the wake, the only solution is to run a full self-consistent simulation. In the case of beam loading, the transfer matrix approach is still useful because first, the leading particle sheet of the bunch can be modeled accurately, and understanding its behavior should still be useful in design before running a full scale simulation. Second, the fields of an ultrarelativistic beam don't change much with $\gamma$, and the beam loading affects the longitudinal field. The transverse force due to the bunch itself is effectively cancelled for co-moving particles. Moreover, for an ultrarelativistic beam, the lack of dispersion means the bunch shape will stay the same as it is accelerated. This means a single simulation could be run with a witness beam of a given energy (for $\gamma\ggg 1$), and then the fields used to understand the beam transport for stages with different energies / beam profiles under the assumption that the witness beam is injected into the same phase in each stage. Nevertheless, the external field approximation is the most significant limitation of the model.

\section{Acknowledgements}
We acknowledge support from U.S. NSF grant 1804463 and AFOSR grant FA9550-19-1-0072. 	The authors would like to acknowledge the OSIRIS Consortium, consisting of UCLA and IST (Lisbon, Portugal) for providing access to the OSIRIS 4.0 framework. Work supported by NSF ACI-1339893. 
\appendix
\section{Betatron phase error}\label{phaseerror}
To estimate the phase error from the symplectic second order scheme for betatron oscillations, we consider the $2\times2$ submatrix representing motion in the $x$ direction only:
\begin{equation}
    M_x = \begin{bmatrix}
1-\frac{\alpha_{xx}^2\Delta t^2}{\gamma} & \frac{\Delta t}{\gamma} \\
- \alpha_{xx}^2\Delta t & 1 
\end{bmatrix}\;,
\end{equation}
where we have dropped  indices for timestep etc. for clarity.  
This can be compared with the usual transfer matrix solution \cite{Wiedemann_book};
\begin{equation}
\tilde{M}_x=\exp[A\Delta t]\equiv
\begin{bmatrix}
C_x & S_x \\
C_x^\prime & S_x^\prime 
\end{bmatrix}
\end{equation}
where $$C_x=\cos \left(\frac{\alpha_{xx}\Delta t}{\sqrt{\gamma}}\right)\;,$$ $$S_x = \frac{1}{\alpha_{xx}\sqrt{\gamma}}\sin \left(\frac{\alpha_{xx}\Delta t}{\sqrt{\gamma}}\right)$$ and the prime $\prime$ denotes the derivative with proper time $\tau$, since $\tilde{M}_x$ represents solutions to the oscillator equation
\begin{equation}
    \frac{d^2x}{d\tau^2} = -\gamma\alpha^2_{xx} x\;.\label{eqn4}
\end{equation}
As both matrices have a determinant of 1, the eigenvalues $\lambda$ of $Y\in\{M_x,\tilde{M}_x\}$ are given by 
$$
\lambda = \frac{\Tr{Y}}{2} \pm i\sqrt{1 - \left(\frac{\Tr{Y}}{2}\right)^2}\;.
$$
Hence, the exact solution has eigenvalues $\lambda_\pm = \exp(\pm i\alpha_{xx}\Delta t/\sqrt{\gamma})$ 
with phase angle $\theta = \alpha_{xx}\Delta t/\sqrt{\gamma}$, as expected, whereas the second order solution has eigenvalues 
$$
\lambda_\pm = 1-\frac{\alpha_{xx}^2\Delta t^2}{2\gamma} \pm i\sqrt{ 1- \left(1-\frac{\alpha_{xx}^2\Delta t^2}{2\gamma}\right)^2}
$$
with phase angle
$$
\theta = \arctan\left(\frac{\sqrt{ 1- \left(1-\frac{\alpha_{xx}^2\Delta t^2}{2\gamma}\right)^2}}{1-\frac{\alpha_{xx}^2\Delta t^2}{2\gamma}}\right)\;.
$$
Expanded for small $\alpha_{xx}\Delta t/\sqrt{\gamma}$, this can be expressed as
$$
\theta = \frac{\alpha_{xx}\Delta t}{\sqrt{\gamma}} + \frac{1}{24}\left(\frac{\alpha_{xx}\Delta t}{\sqrt{\gamma}}\right)^3+\dots
$$
Hence, the betatron frequency is larger by a factor of $1 + \alpha_{xx}^2\Delta t^2/(24\gamma)$ using the symplectic second order scheme compared with the analytic solution. Since in numerical simulations to resolve the plasma dynamics $\alpha_{xx}^2\Delta t^2\lesssim1$, i.e. $\alpha_{xx}^2\Delta t^2/(24\gamma)\lll 1$, this is generally a negligible error.

\section{Nonlinear transfer matrix: Explicit forms, validation and accuracy of solutions}\label{nonlindetails}
In this appendix we give  explicitly forms of the matrix $\mathcal{M}$ for the purposes of clarity. For the simpler system in phasespace coordinates $x,u_x$ only, the transfer matrices are 
\begin{equation}
    M = \begin{bmatrix}
1-\frac{\alpha^2_{xx}\Delta t^2}{\gamma} & \frac{\Delta t}{\gamma} \\
- \alpha^2_{xx}\Delta t & 1 
\end{bmatrix}\;,\quad {M_D} = \begin{bmatrix}
-\frac{\alpha^2_{xx}\Delta t^2}{\gamma} & \frac{\Delta t}{\gamma} \\
0 & 0 
\end{bmatrix}\;,
\end{equation}
but it is straightforward to extend this analysis to the $4\times 4$ transfer matrix. First, we expand  to first order in $\delta\gamma$ only. The particle coordinates including the nonlinear terms are
\begin{equation}
    w_\delta = \begin{bmatrix} 1\\\delta\gamma \end{bmatrix} \otimes w = \begin{bmatrix} x\\u_x\\\delta\gamma x \\ \delta\gamma u_x\end{bmatrix}
\end{equation}
and the corresponding  transfer matrix is, in block matrix and explicit forms respectively, 
\begin{equation}
    \mathcal{M} =\begin{bmatrix} M & -\tfrac{1}{\gamma}{M_D}\\ 0& M \end{bmatrix} = \begin{bmatrix}
1-\frac{\alpha^2_{xx}\Delta t^2}{\gamma} & \frac{\Delta t}{\gamma} &\frac{\alpha^2_{xx}\Delta t^2}{\gamma^2} & -\frac{\Delta t}{\gamma^2}\\
- \alpha^2_{xx}\Delta t & 1 & 0 & 0 \\ 
0 & 0 & 1-\frac{\alpha^2_{xx}\Delta t}{\gamma} & \frac{\Delta t}{\gamma}\\
0 & 0 & - \alpha^2_{xx}\Delta t &1
\end{bmatrix}\;.
\end{equation}

We may extend this process to any order in $\delta\gamma$, for example for expansion in a series beyond 6th order, the matrix in block matrix form is
\begin{widetext}
\begin{equation}
    \mathcal{M} =\begin{bmatrix} M & -\tfrac{1}{\gamma}M_D & \tfrac{1}{\gamma^2}M_D& -\tfrac{1}{\gamma^3}M_D& \tfrac{1}{\gamma^4}M_D & -\tfrac{1}{\gamma^5}M_D& \tfrac{1}{\gamma^6}M_D &\dots\\ 
    0& M& -\tfrac{1}{\gamma}M_D & \tfrac{1}{\gamma^2}M_D& -\tfrac{1}{\gamma^3}M_D & \tfrac{1}{\gamma^4}M_D & -\tfrac{1}{\gamma^5}M_D&\dots\\
    0&0&M& -\tfrac{1}{\gamma}M_D & \tfrac{1}{\gamma^2}M_D& -\tfrac{1}{\gamma^3}M_D & \tfrac{1}{\gamma^4}M_D &\dots\\
    0&0&0&M& -\tfrac{1}{\gamma}M_D & \tfrac{1}{\gamma^2}M_D& -\tfrac{1}{\gamma^3}M_D & \\
    0&0&0&0& M& -\tfrac{1}{\gamma}M_D & \tfrac{1}{\gamma^2}M_D &\dots\\
    0&0&0&0&0& M& -\tfrac{1}{\gamma}M_D&\dots \\
    0&0&0&0&0&0& M&\dots\\
    \vdots&\vdots&\vdots&\vdots&\vdots&\vdots&\vdots
    \end{bmatrix} \;.\label{eq:bigmatrix}
    \end{equation}
\end{widetext}

The reason for expanding to high order is to allow a large phase difference due to energy spread to accumulate without error. The expansion means that the transfer matrix is no longer symplectic. We need a way of calculating how many terms are needed in this expansion for a given situation. Although there may be in general a complicated field variation, we can estimate the betatron phase accumulated for a given situation and use this to estimate the number of terms needed in the expansion. 

For the nonlinear matrix expanded in $\delta\gamma^m$ described above, the maximum term required in $\delta\gamma^m$ can be estimated through expansion of the eigenmodes. Assuming the time-step is small, $\Delta t\rightarrow 0$, the eigenvalues of $M$ approach $$\exp\left(\pm i \int_0^t\frac{\alpha_{xx}(t^\prime) dt^\prime}{\sqrt{\gamma(t^\prime)}}\right)$$ (for discussion on the finite difference phase error, refer to section \ref{phaseerror}). For a particle with (normalized) energy $\gamma +\delta\gamma$,
the eigenvalues will be $\exp(\pm i \int_0^t\alpha_{xx}(t^\prime) dt^\prime/\sqrt{\gamma(t^\prime)+\delta\gamma})$. Writing the phase for $\delta\gamma = 0$ as $\psi_0 = \int_0^t\alpha_{xx}(t^\prime) dt^\prime/\sqrt{\gamma(t^\prime)}$, consider first that the phase for a particle with an energy deviating by $\delta\gamma$ can be expanded as
$$
\exp\left[\pm i\psi_0\pm i\int_0^{\psi_0}d\psi\left(-\frac{1}{2}\frac{\delta\gamma}{\gamma} +\frac{3}{4}\frac{\delta\gamma^2}{\gamma^2}+\dots\right)\right] \;,
$$
where $d\psi \equiv \alpha_{xx}(t^\prime) dt^\prime/\sqrt{\gamma(t^\prime)}$. 
After factoring out $\exp(i\psi_0)$, expanding the remaining exponential term will result in many terms in higher powers of $\delta\gamma/\gamma$. Since we are interested in considering large phase $\psi_0$ (i.e., many betatron oscillations), however, the magnitude of the largest term at any order $m$ in $\delta\gamma/\gamma$ will in general be $|\int_0^{\psi_0}d\psi\delta\gamma/\gamma|^m/2m!$ and hence, to determine how many orders are needed for an accurate solution, we require
\begin{equation}
\left|\frac{\delta\gamma}{2}\int_0^{\psi_0}\frac{d\psi}{\gamma}\right|^m\ll m!\label{acc_check}
\end{equation} 
for the highest order $m$ in the expansion.

\begin{figure}[htbp]
\begin{center}
\includegraphics[width =0.5\textwidth]{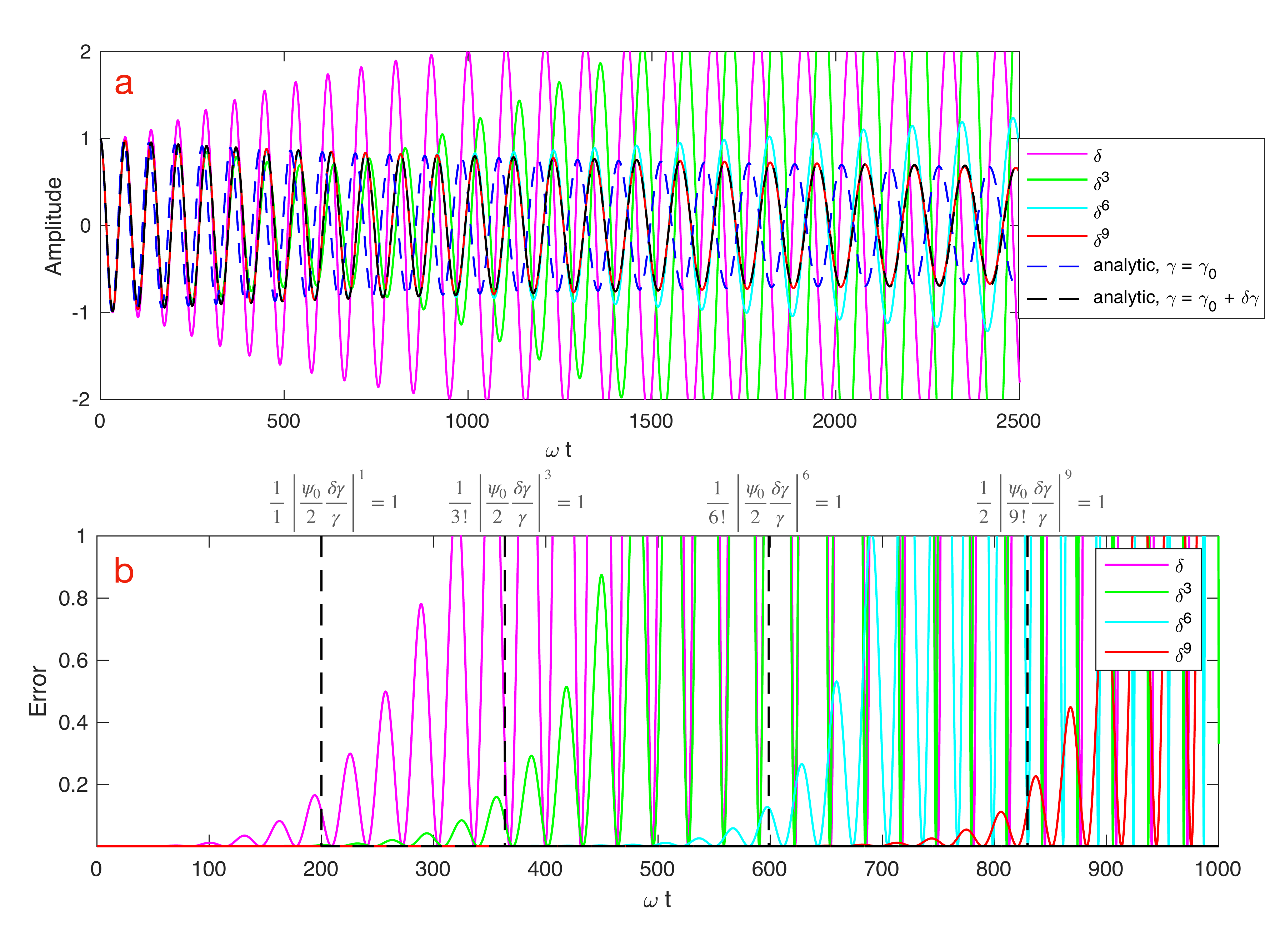}
\caption{Betatron oscillations modelled using the expanded nonlinear transfer matrix in Eqn.~\eqref{eq:bigmatrix}. (a) Oscillations with an increasing $\gamma$ factor from $\gamma_0 = 100$ to $\gamma_0 = 500$ corresponding to linear acceleration. The dashed lines correspond to analytic (WKB) solutions for betatron oscillations for the design $\gamma = \gamma_0$ (blue) and $\gamma = \gamma_0+\delta\gamma$ (black), for a energy deviation $\delta\gamma = 10$. The colored solid lines indicate solutions using the nonlinear matrix with different orders in $\delta = \delta\gamma / \gamma$ up to the $\delta^9$th term, i.e. $m=9$. (b) The error in the solution for oscillations with fixed $\gamma = 100$, defined as $|x_{n\delta^m} - x|^2$, where $x$ is the analytic solution and $x_{n\delta^m}$ is the matrix solution including terms up to $m$. The black dashed lines show the thresholds $\left|\frac{\psi_0}{2}\frac{\delta\gamma}{\gamma}\right|^m/ m! = 1$ corresponding to Eqn.~\eqref{acc_check}.  }
\label{fig:bigmatrix}
\end{center}
\end{figure}

Fig.~\ref{fig:bigmatrix} shows betatron oscillations modelled using the expanded nonlinear transfer matrix in Eqn.~\eqref{eq:bigmatrix}. Fig.~\ref{fig:bigmatrix}a shows oscillations with an increasing $\gamma$ factor from $\gamma_0 = 100$ to $\gamma_0 = 500$ corresponding to linear acceleration. The dashed lines correspond to analytic (WKB) solutions for betatron oscillations for the design $\gamma = \gamma_0$ (blue) and $\gamma = \gamma_0+\delta\gamma$ (black), for a energy deviation $\delta\gamma = 10$. This corresponds to $\delta\gamma/\gamma = 0.1$ initially, which is far larger than any real design, but is chosen to stretch the limits of the approximation. The colored solid lines indicate solutions using the nonlinear matrix with different orders in $\delta = \delta\gamma / \gamma$ up to the $\delta^9$th term, i.e. $m=9$. (b) The error in the solution for oscillations with fixed $\gamma = 100$, defined as $|x_{n\delta^m} - x|^2$, where $x$ is the analytic solution and $x_{n\delta^m}$ is the matrix solution including terms up to $m$. The black dashed lines show the thresholds $\left|\frac{\psi_0}{2}\frac{\delta\gamma}{\gamma}\right|^m/ m! = 1$ corresponding to Eqn.~\eqref{acc_check}. These indicate that for betatron phases less than the threshold phase given by this condition, the error in the matrix solution remains small.

\begin{figure}[htbp]
\begin{center}
\includegraphics[width =0.5\textwidth]{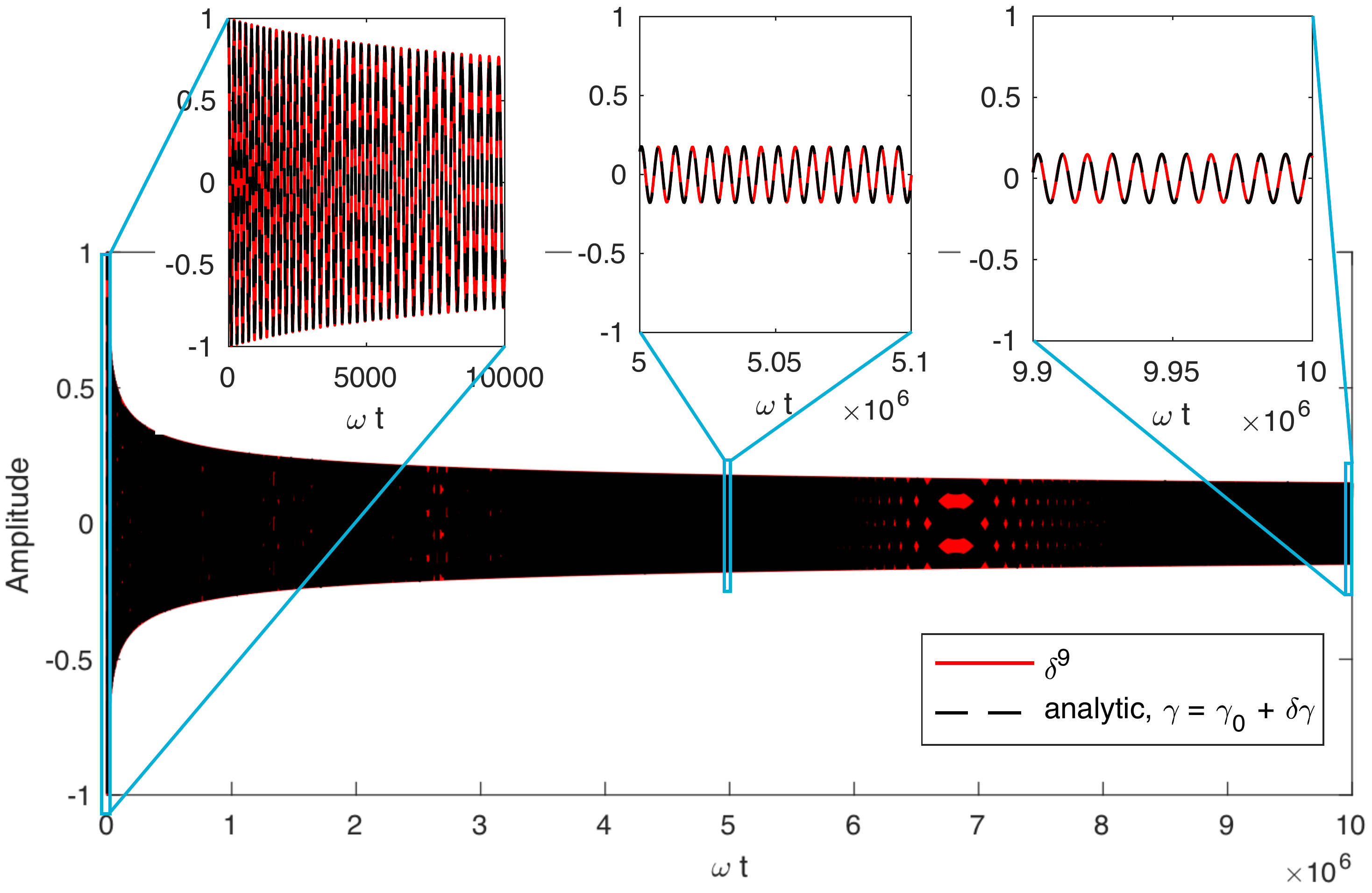}
\caption{Demonstration of method accuracy; Betatron oscillations for particle linearly accelerated with design energy $\gamma = 1000$ to $2\times10^6$ over a total time of $\omega t = 10^7$, where the betatron frequency is $\omega/\sqrt{\gamma}$, with a deviation from the design energy of $\delta\gamma = 20$, i.e. corresponding to an initial relative energy spread $\delta\gamma/\gamma = 2\%$. The main panel shows the (unresolvable) oscillations for a WKB solution compared with the nonlinear matrix solution including terms up to $m=9$ over the full range. The inset panels show expanded regions at the beginning, middle and end.}
\label{fig:TeVsimplebetatron}
\end{center}
\end{figure}

Fig.~\ref{fig:TeVsimplebetatron} shows betatron oscillations for particle linearly accelerated with a design energy from $\gamma = 1000$ to $2\times10^6$ over a total time of $\omega t = 10^7$, where the betatron frequency is $\omega/\sqrt{\gamma}$ and $\omega$ is a constant. The particle has a deviation from the design energy of $\delta\gamma = 20$. These parameters roughly correspond to a particle being accelerated from 500 MeV to 1 TeV in a 100 m long plasma accelerator, for a particle with normalized energy $\gamma+\delta\gamma$ in a beam that has an initial relative energy spread $\delta\gamma/\gamma = 2\%$.  The main panel shows the (unresolvable) oscillations for a WKB solution compared with the nonlinear matrix solution including terms up to $m=9$ over the full range. The inset panels show expanded regions at the beginning, middle and end. These show that the methods described in this manuscript of the nonlinear matrix expanded to 9 orders in $\delta \gamma / \gamma$ can accurately capture the betatron oscillations with negligible phase and zero amplitude error over the full range of acceleration.
{
\section{Scaling in the number of operations for the different methods discussed}
Here, we discuss the number of operations required in the linear and nonlinear transfer matrix methods compared with an imagined generic second-order particle tracking code. $M$ is the basic transfer matrix and $w$ is the system vector. An imagined tracking code would involve a numerical scheme that would be equivalent to the repeated application of $M$ to $w$. For $M$ being represented by an $n\times n$ matrix, at most $n^2$ multiplication and addition operations would be needed per time-step, equivalent to the matrix multiplication $Mw$. For a calculation of $N_t$ timesteps, the number of operations required to calculate the particle trajectory in a tracking code would scale as $\mathcal{O}(N_tn^2)$. Using the linear transfer matrix for a single particle, the number of operations would scale as, at most, $\mathcal{O}(N_tn^3)$, with the extra factor of $n$ because it involves repeated matrix multiplication $M M$ rather than $M w$. 

For a large number of particles, $N_p$, being tracked through the same field structure and with the same energy, the number of operations needed for calculating the end state of the particles scales as $\mathcal{O}(N_pN_tn^2)$ for a tracking code, but ${\mathcal{O}(N_{\max(t,p)}n^3)}$, where $N_{\max(t,p)} = \max(N_p,N_t)$, for the transfer matrix method as the full transfer matrix only needs calculating once. Hence, $\mathcal{O}(N_p/n)$ more operations are required for a particle tracing code to calculate the final phase-space positions of $N_p$ particles. For $N_p\gg n$, this is evidently a substantial computational saving ($n$ is either 2 or 4, but $N_p$ may be $10^4$ or more for good statistics). Moreover, the matrix method can be used to transform the beam phase space rather than individual particle tracks.

For the $(m+1)n\times (m+1)n$ nonlinear matrix, $\mathcal{M}$, the number of operations in the calculation $\mathcal{M}\mathcal{M}$ scales as (at most) $\mathcal{O}(m^3n^3)$, which means that the method is (at most) $\mathcal{O}(m^3n)$ times more expensive than using a simple particle tracing code for a single trajectory. This appears to be undesirable as for $m=9$, there are $\mathcal{O}(10^3)$ more operations required, which even for a very large number of particle tracks is not a favorable scaling. However, if the phase-spaces of particles with a number of different energies, $N_\gamma$, are of interest, as in the study in this manuscript, then even the linear matrix method would need to be calculated for each energy, so the number of operations would be $\mathcal{O}(N_\gamma N_{\max(t,p)}n^3)$. Whereas, for the nonlinear matrix, the number of operations required would be $\mathcal{O}(N_{\max(t,p)}m^3n^3)$ (as $N_\gamma\leq N_p$). This means that the linear matrix method would require $\mathcal{O}(N_\gamma/m^3)$ more operations than the nonlinear method. For random sampling of energies to generate a gaussian distribution, $N_\gamma = N_p$, as each particle requires a randomly sampled energy. Therefore this can be a significant saving if $N_p\gg m^3$, as in our study ($N_p=10^5$, $m^3 \sim 10^3$).

For the $N = 200\times200$ point parameter space of particle phase spaces we investigated in Figs.~\ref{fig:emitt_grow_final} and \ref{fig:emitt_grow_dgog}, the number of operations required for a generic tracking code would scale as $\mathcal{O}(N N_p N_t n^2)$ compared with $\mathcal{O}(N_tm^3n^3)$ for the nonlinear matrix method ($N_t\gg N_p$ in our studies), i.e., the full calculation of $N_p$ particles needs repeating $N$ times for the tracing code, but once the nonlinear matrix for the lattice is generated, the parameter space is investigated with $N$ operations of size $m^3n^3$ only (this assumes that $N<N_{\max(t,p)}$). Hence, the ratio of the number of operations required for a generic tracking code compared to the nonlinear matrix method for the parameter space investigated here scales as $\mathcal{O}(N N_p / m^3n)$. With $N =4\times10^4$, $N_p=10^5$, $n = 2$ and $m=9$, this ratio is $\mathcal{O}(10^6)$, which would have made the total calculation for this paper that ran in 143 seconds (on a 4 GHz Intel Core i7 Macintosh computer, not including the particle-in-cell calculation of the field structure, vectorized code written in MATLAB 2020a) unfeasible without making use of a large computing cluster. 
}

\end{document}